\documentclass[letterpaper, 10 pt, conference]{ieeeconf}  

\IEEEoverridecommandlockouts                              

\usepackage{import}
\usepackage{cite}
\usepackage{graphics} 
\usepackage{epsfig} 
\usepackage{times} 
\usepackage{amsmath} 
\usepackage{amssymb}  
\usepackage{lipsum}
\usepackage{multicol}
\usepackage{graphicx}
\usepackage{afterpage}
\usepackage{subcaption}
\usepackage{multirow}
\usepackage{layout}
\usepackage{gensymb}
\usepackage{booktabs}

\DeclareMathOperator*{\argmin}{argmin}   

\newcommand{\confidentialheader}{FOR REVIEW ONLY}
\newcommand{\conferencesubmission}{Preprint submitted to the 2018 IEEE International Conference on Robotics and Automation.}

\usepackage{fancyhdr}
\fancypagestyle{pageStyleOne}{%
    \fancyhead[C]{\confidentialheader}
    \fancyfoot[C]{\conferencesubmission}
}


\title{\LARGE \bf
Multi-Hypothesis Visual-Inertial Flow 
}

\author{E. Jared Shamwell$^{1}$, William D. Nothwang$^{2}$, Donald Perlis$^{3}$
\thanks{*This work was supported by the US Army Research Laboratory}
\thanks{$^{1}$E. Jared Shamwell, PhD is a research engineer with GTS stationed at the US Army Research Laboratory, Adelphi, MD 20783.
        {\tt\small earl.j.shamwell.ctr@mail.mil; ejsham@umd.edu}}%
\thanks{$^{2}$William D. Nothwang, PhD is the Branch Chief (a) of the Micro and Nano Devices and Materials Branch in the Sensors and Electron Devices Directorate at the US Army Research Laboratory, Adelphi, MD 20783.
        {\tt\small william.d.nothwang.civ@mail.mil }}%
\thanks{$^{3}$Donald Perlis, PhD is a Professor of Computer Science at the University of Maryland, College Park, MD 20742.
        {\tt\small perlis@umd.edu}}%
}

\begin{document}

\maketitle

\begin{abstract}

Estimating the correspondences between pixels in sequences of images is a critical first step for a myriad of tasks including vision-aided navigation (e.g., visual odometry (VO), visual-inertial odometry (VIO), and visual simultaneous localization and mapping (VSLAM)) and anomaly detection. We introduce a new unsupervised deep neural network architecture called the Visual Inertial Flow (VIFlow) network and demonstrate image correspondence and optical flow estimation by an unsupervised multi-hypothesis deep neural network receiving grayscale imagery and extra-visual inertial measurements. VIFlow learns to combine heterogeneous sensor streams and sample from an unknown, un-parametrized noise distribution to generate several ($4$ or $8$ in this work) probable hypotheses on the pixel-level correspondence mappings between a source image and a target image. We quantitatively benchmark VIFlow against several leading vision-only dense correspondence and flow methods and show a substantial decrease in runtime and increase in efficiency compared to all methods with similar performance to state-of-the-art (SOA) dense correspondence matching approaches. We also present qualitative results showing how VIFlow can be used for detecting anomalous independent motion.

\end{abstract}


\section{Introduction}

State estimation for size, weight, power, and computation (SWaP-C) constrained robotic systems is limited by the lightweight and low-power sensing and computational hardware that they are forced to use. When viewed as complex, embodied agents, robotic systems can generate and access a wide variety of varied sensory information, and as such, a popular approach to mitigating the negative influences of noisy SWaP-C constrained sensors is to fuse estimates from an array of heterogeneous sensors deployed on the robot.

\begin{figure}[th]
\centering
	\begin{minipage}{.45\textwidth}
		\begin{subfigure}{\linewidth}
    			\begin{subfigure}{0.24\linewidth}
  	   			\includegraphics[width=\linewidth]{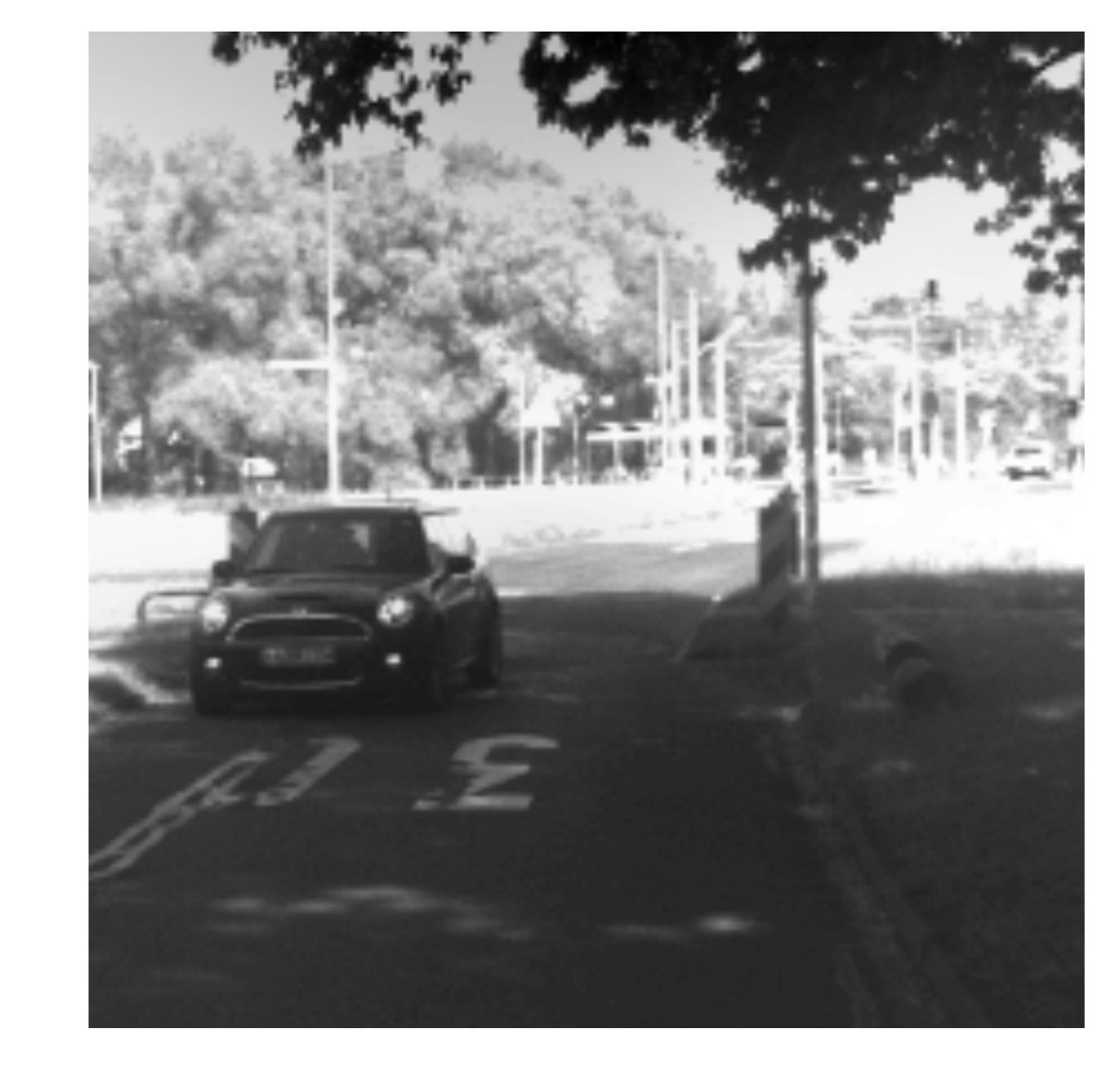}
  	 		\end{subfigure}%
    			\begin{subfigure}{0.24\linewidth}
  	   			\includegraphics[width=\linewidth]{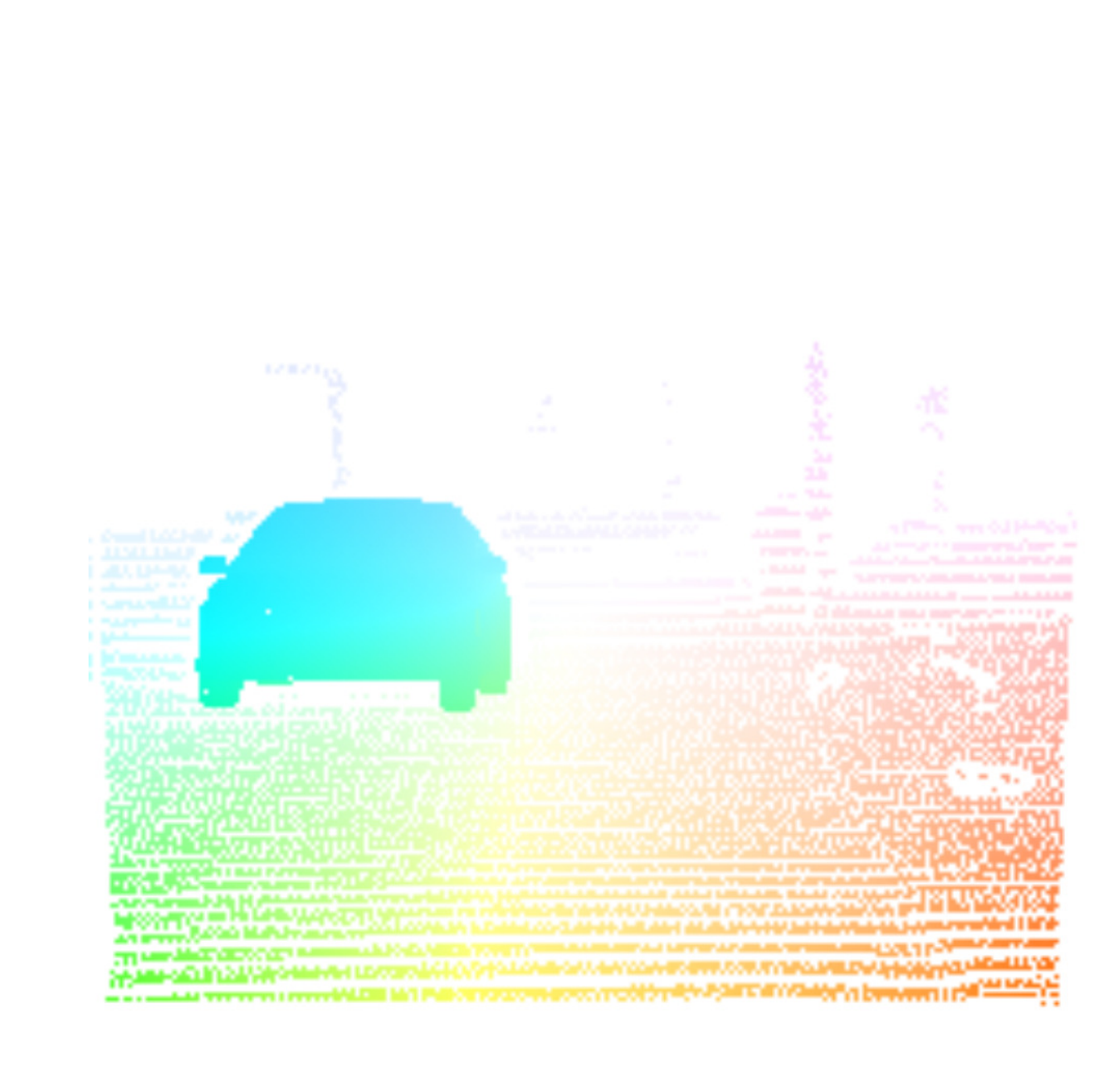}
  	 		\end{subfigure} 
    			\begin{subfigure}{0.24\linewidth}
  	   			\includegraphics[width=\linewidth]{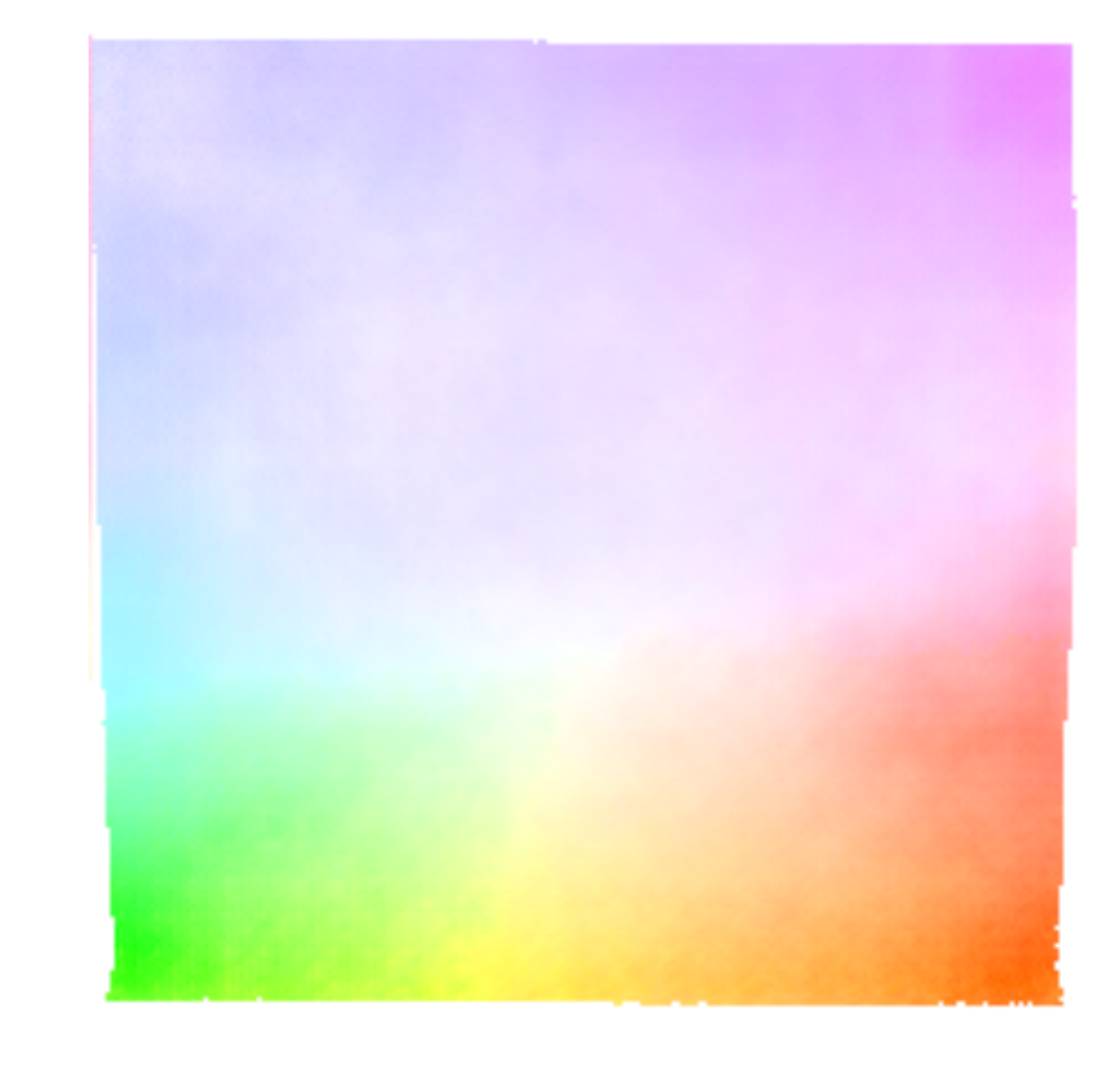}
  	 		\end{subfigure}%
    			\begin{subfigure}{0.24\linewidth}
  	   			\includegraphics[width=\linewidth]{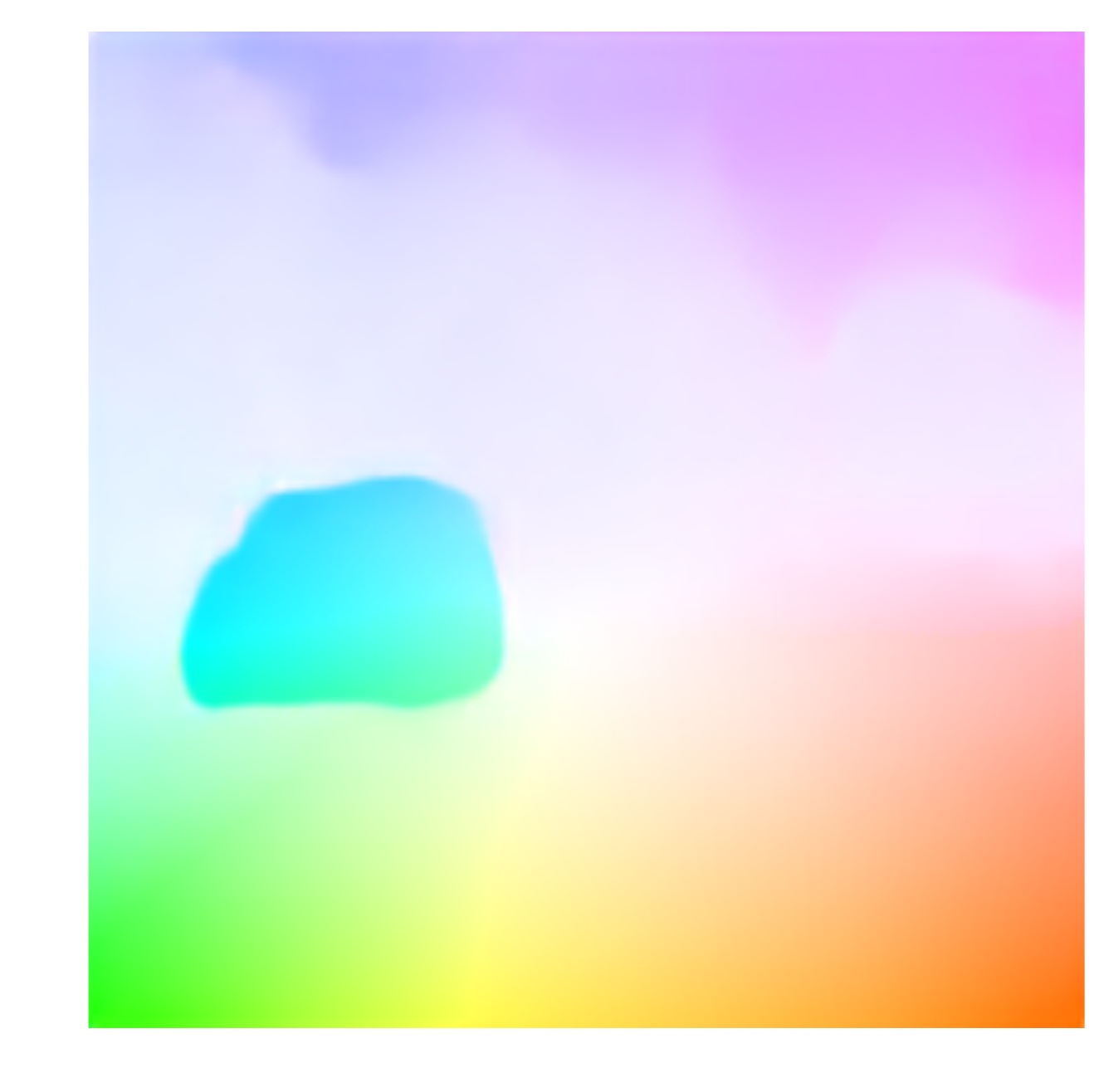}
  	 		\end{subfigure}%
	 	 \end{subfigure}%
	 \end{minipage}
	\begin{minipage}{.45\textwidth}
		\begin{subfigure}{\linewidth}
    			\begin{subfigure}{0.24\linewidth}
  	   			\includegraphics[width=\linewidth]{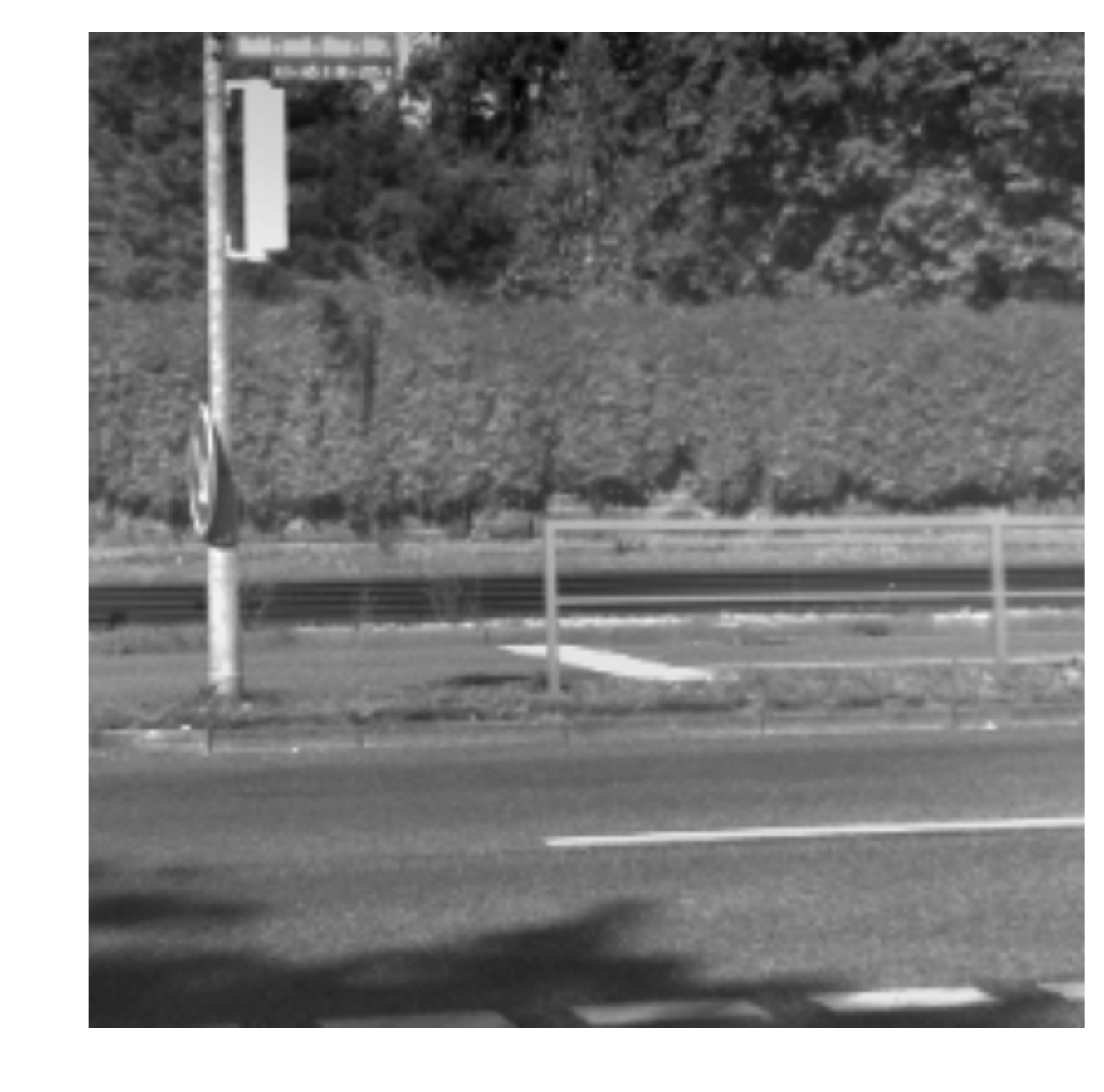}
  	 		\end{subfigure}%
    			\begin{subfigure}{0.24\linewidth}
  	   			\includegraphics[width=\linewidth]{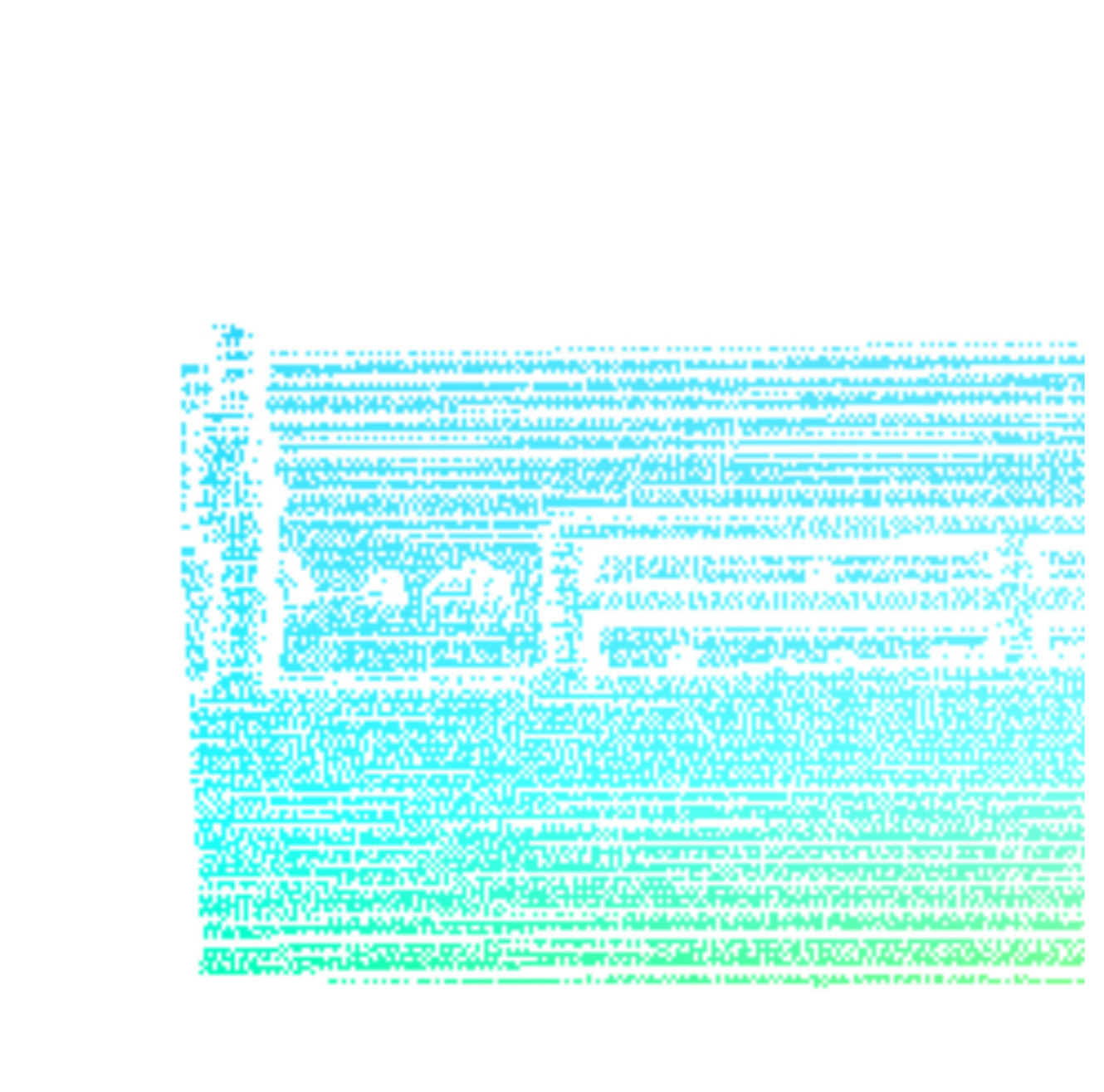}
  	 		\end{subfigure} 
    			\begin{subfigure}{0.24\linewidth}
  	   			\includegraphics[width=\linewidth]{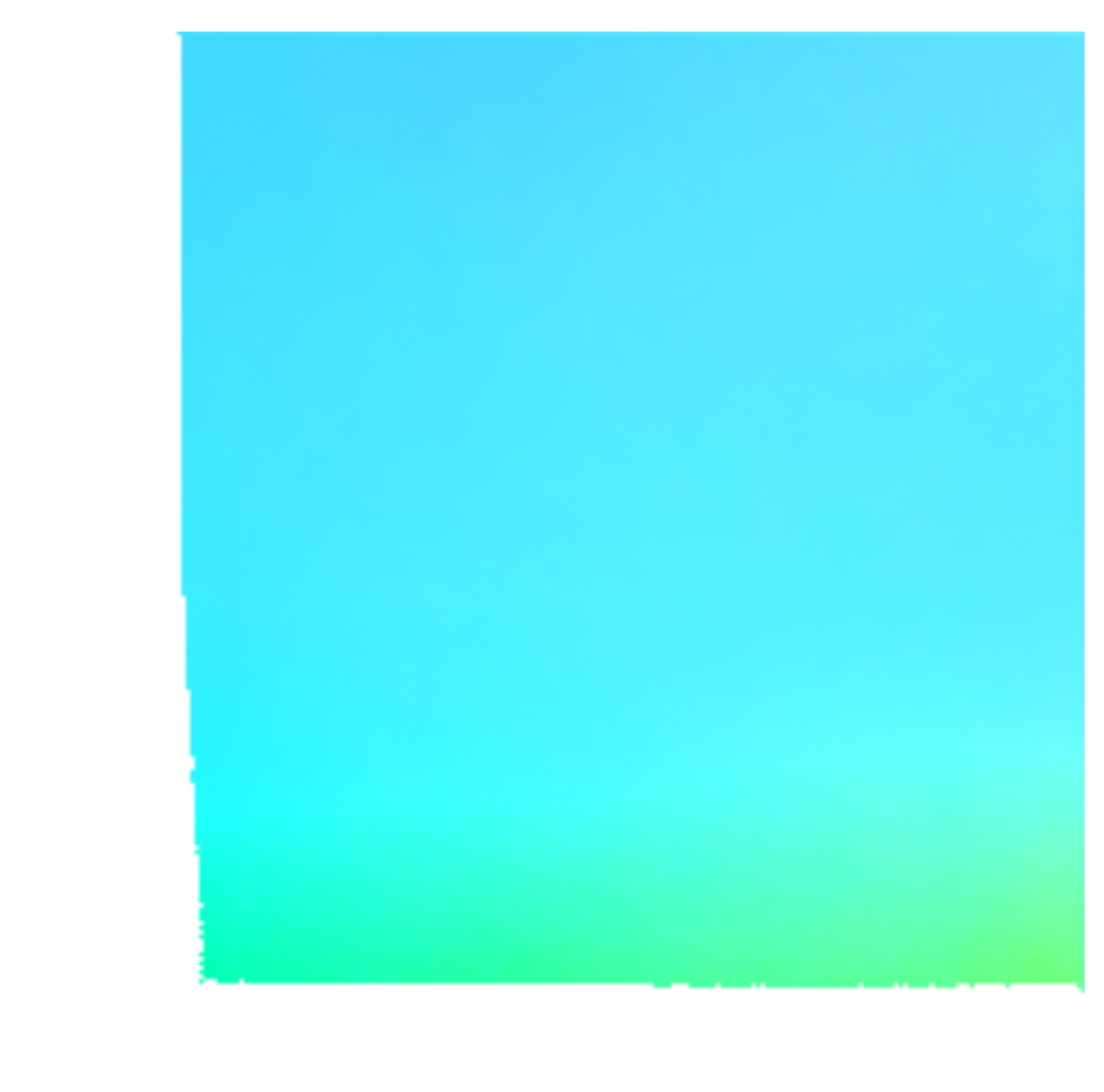}
  	 		\end{subfigure}%
    			\begin{subfigure}{0.24\linewidth}
  	   			\includegraphics[width=\linewidth]{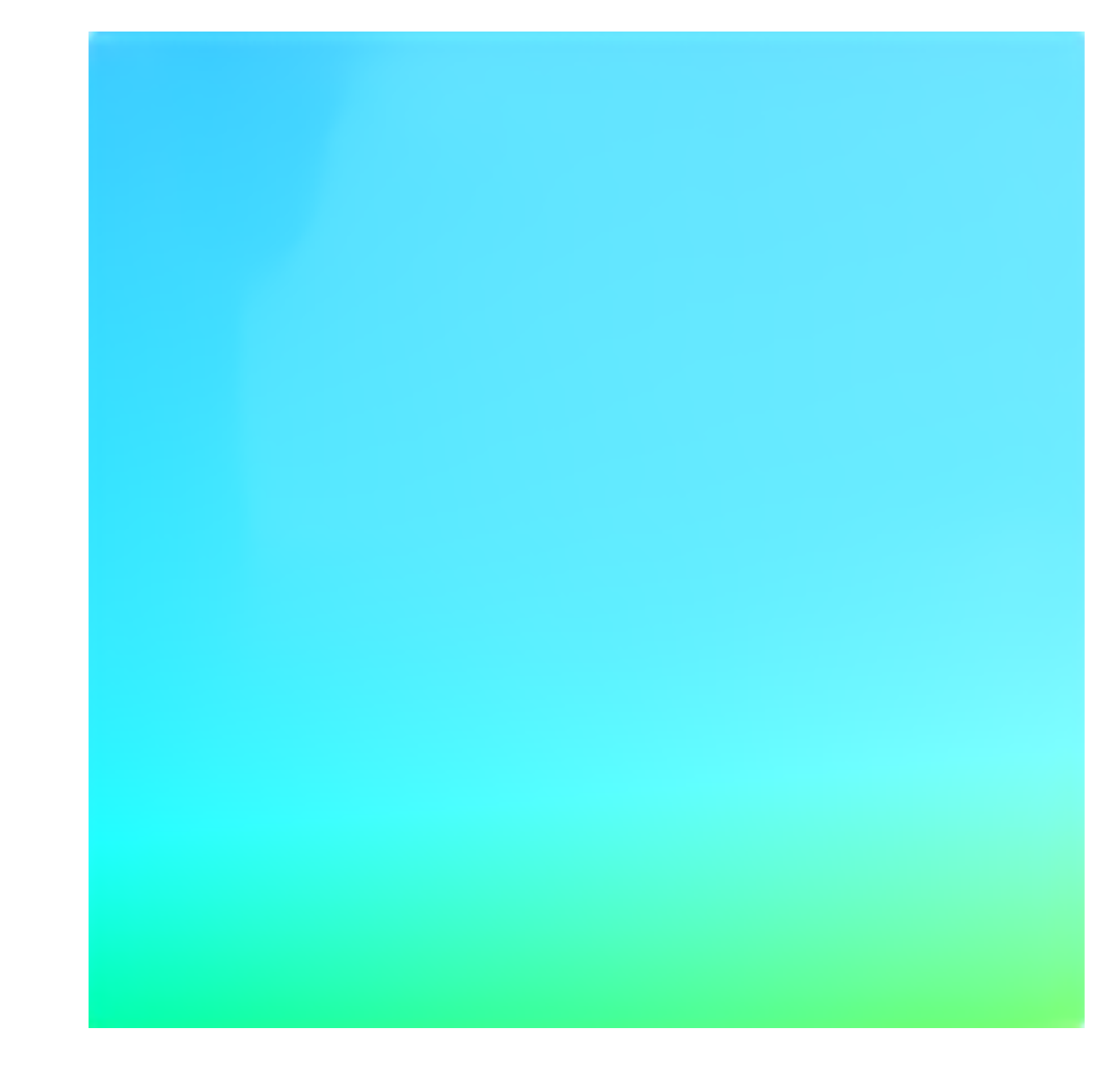}
  	 		\end{subfigure}%
	 	 \end{subfigure}%
	 \end{minipage}	 
	\begin{minipage}{.45\textwidth}
		\begin{subfigure}{\linewidth}
    			\begin{subfigure}{0.24\linewidth}
  	   			\includegraphics[width=\linewidth]{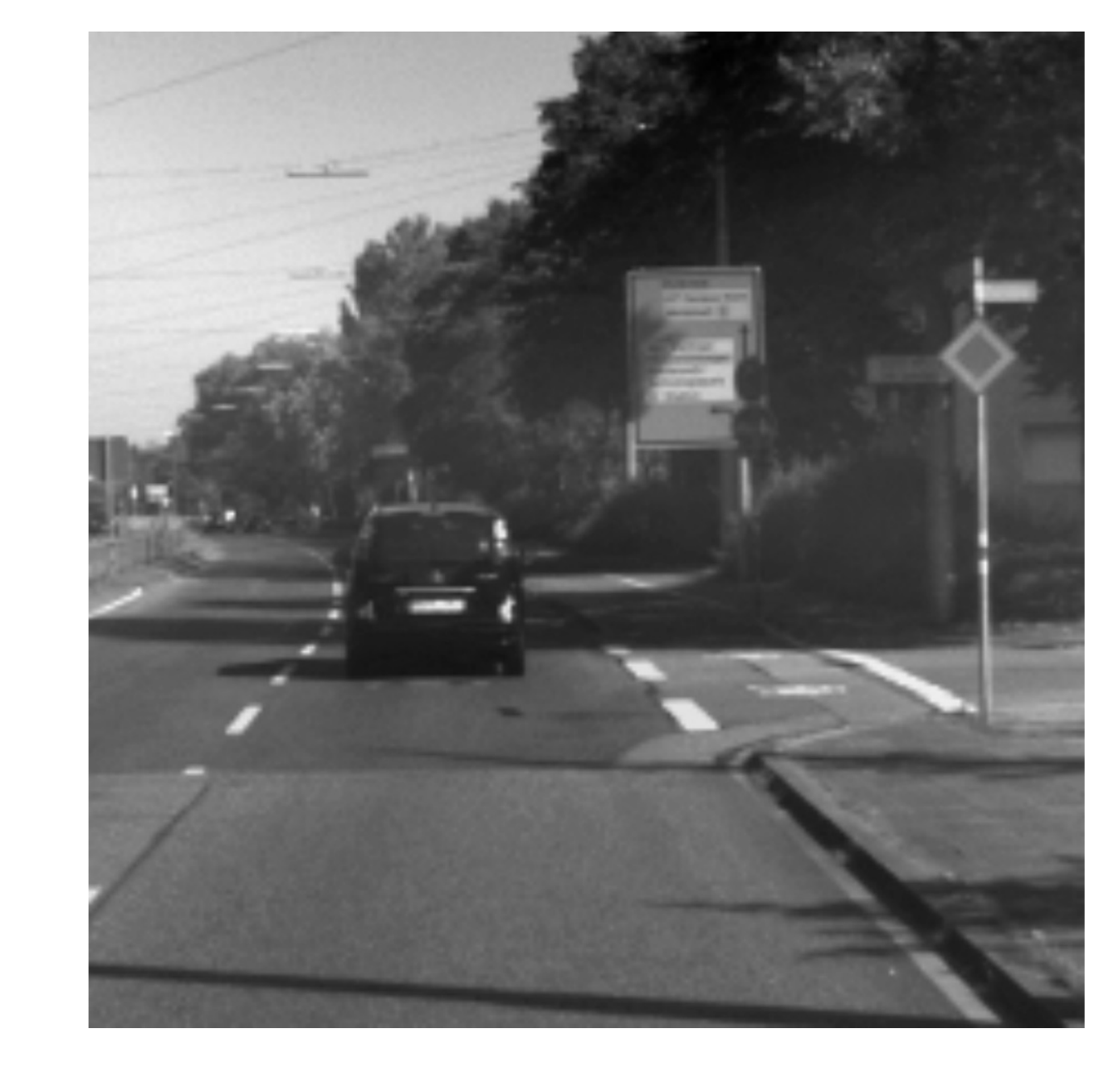}
  	   			\caption{Source}
  	 		\end{subfigure}%
    			\begin{subfigure}{0.24\linewidth}
  	   			\includegraphics[width=\linewidth]{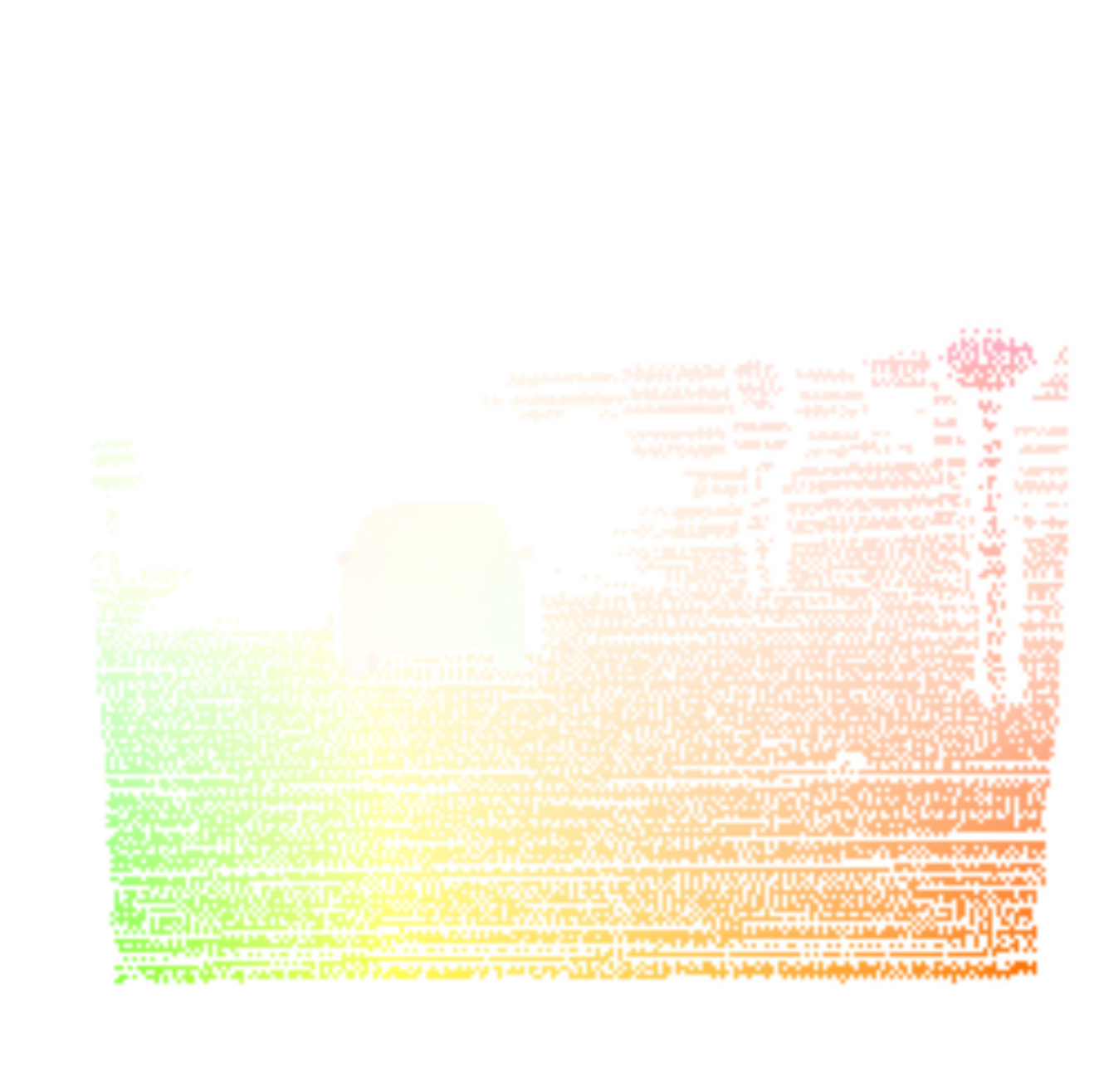}
  	   			\caption{GT flow}
  	 		\end{subfigure} 
    			\begin{subfigure}{0.24\linewidth}
  	   			\includegraphics[width=\linewidth]{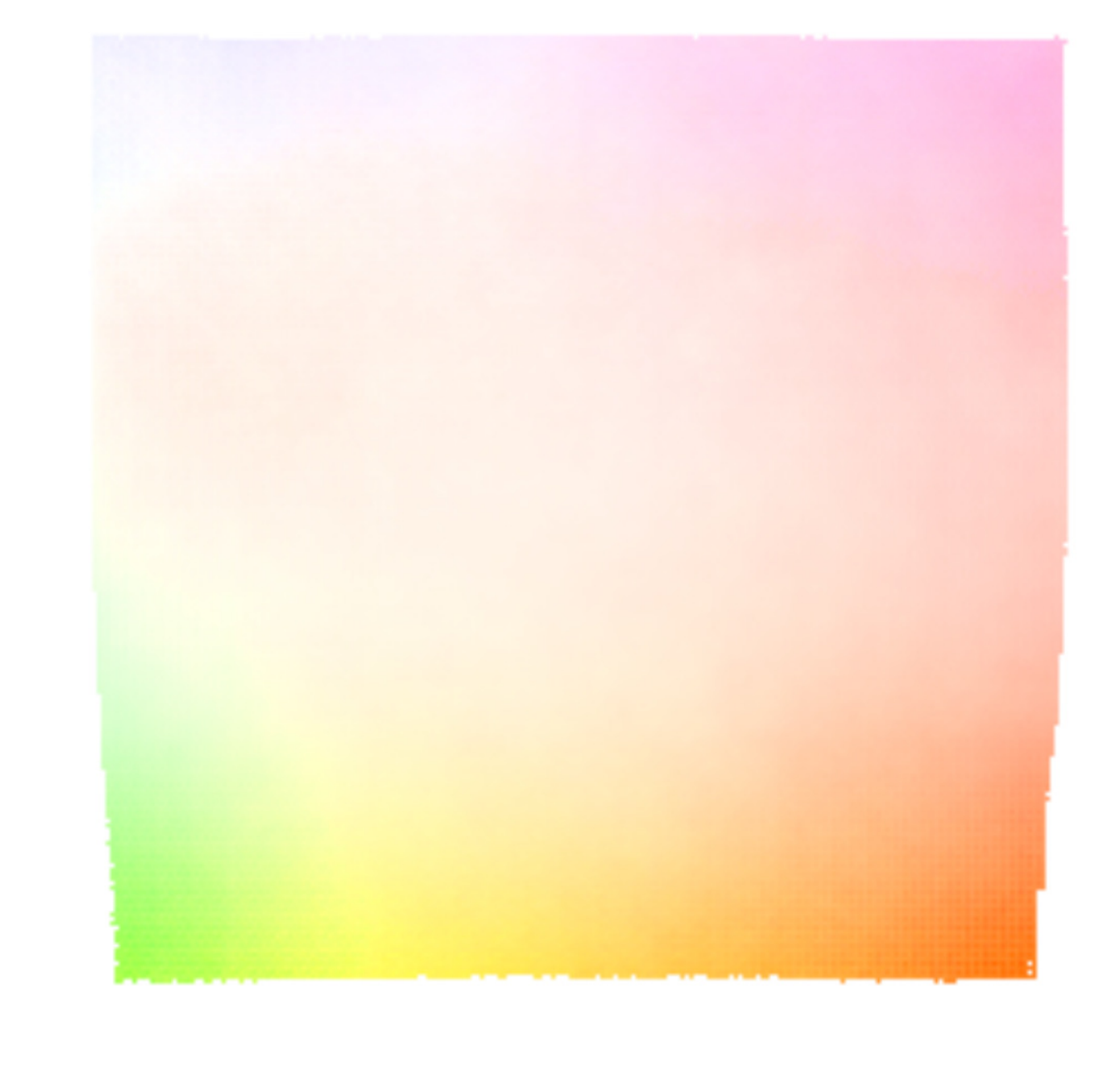}
  	   			\caption{VIFlow}
  	 		\end{subfigure}%
    			\begin{subfigure}{0.24\linewidth}
  	   			\includegraphics[width=\linewidth]{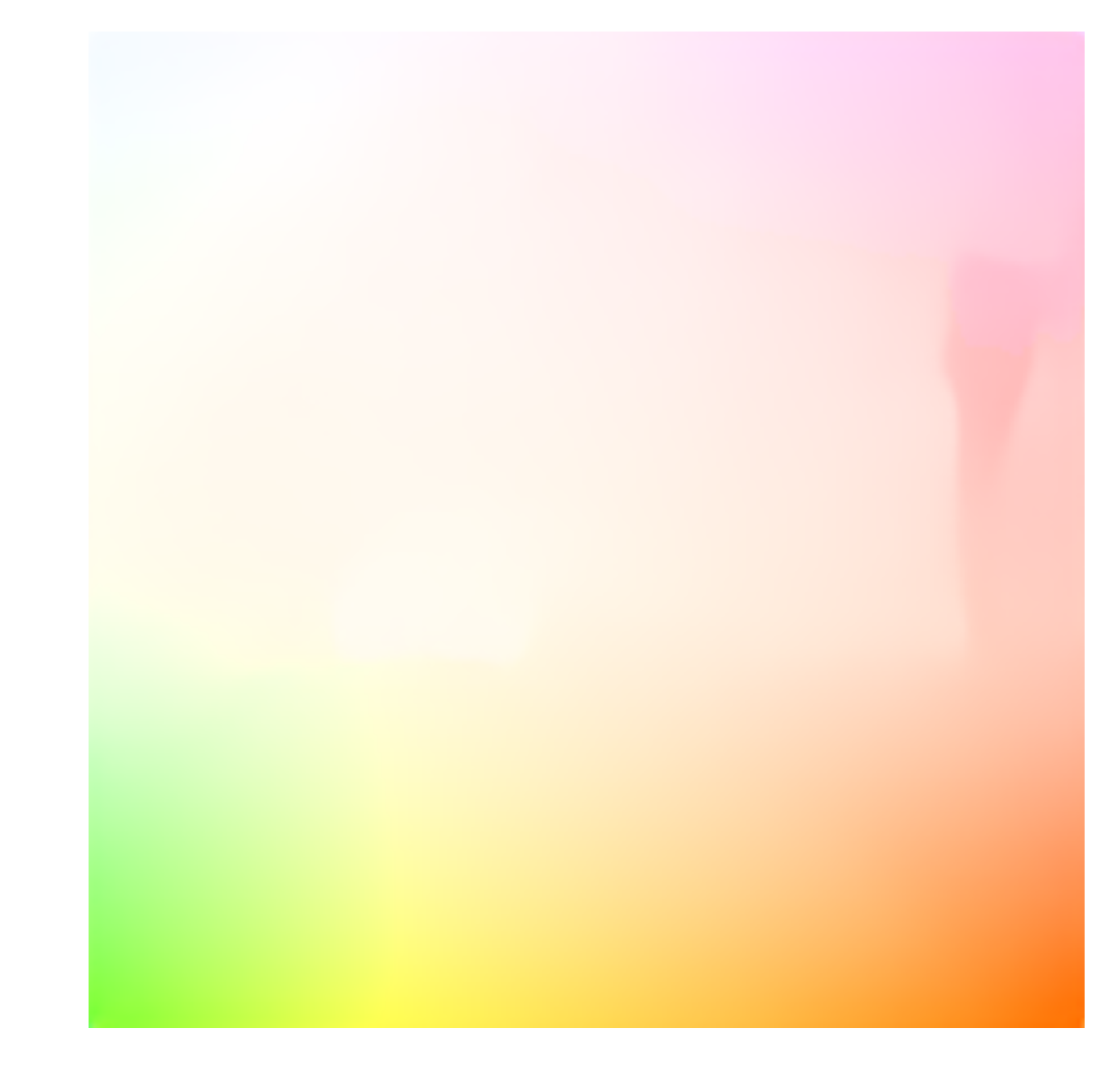}
  	   			\caption{Flownet2}
  	 		\end{subfigure}%
	 	 \end{subfigure}%
	 \end{minipage}		 
	\caption{Sample KITTI \cite{Geiger2013} visual flow. The source image, ground-truth flow, flow from our VIFlow, and flow from Flownet2 \cite{ilg2017} are shown. }		
	\label{sample_flow}	
\end{figure}

SWaP-C constrained GPS-denied navigation has been greatly influenced by this  philosophy of sensor fusion and approaches in visual-inertial odometry (VIO), where a sensor array will commonly consist of a camera and an inertial measurement unit (IMU), have been particularly successful for GPS-denied SWaP-C constrained localization and navigation. 

Such vision-aided approaches to localization generally require finding a correspondence mapping between scene elements in sequential image frames.  To-date, solutions to the correspondence problem (see \cite{scharstein2002} for a review) remain dominated by bottom-up, vision-only approaches that neglect other available sources of complementary information.

We propose to fully exploit the embodied nature of robotic systems and begin fusing heterogeneous sensor measurements as early as possible by learning a multi-hypothesis feed-forward model that receives heterogeneous information and estimates dense image correspondences and flow so as to avoid the computationally heavy direct image matching and subsequent optimization and outlier rejection of most current approaches (see ÒSection \ref{related_work}Ò for more detail).

The main contribution of this paper is an end-to-end trainable, unsupervised deep neural network architecture that:
\begin{enumerate}
\item{Learns to estimate dense visual correspondence/flow using extra-visual sensory information}
\item{Runs substantially faster and more efficiently than state-of-the-art (SOA), vision-only approaches}
\item{Requires no a priori calibration or information about sensor streams}
\item{Learns to estimate anomalous independent motion}
\item{Only requires information freely available on-board a robotic system for training.}
\end{enumerate}

The remainder of the paper is organized as follows: ÒSection \ref{related_work}Ò presents related work; ÒSection \ref{approach}Ò outlines our deep network approach to fusing noisy heterogeneous sensory inputs and describes the network architecture; ÒSection \ref{methods}Ò describes our experimental approach; ÒSection \ref{evaluation}Ò describes our evaluation procedures; ÒSection \ref{results}Ò presents and discusses our experimental results; and ÒSection \ref{conclusion}Ò offers concluding thoughts, limitations, and directions for future work.

\section{Related Work} \label{related_work}

\begin{figure*}[th]
\centering
	\includegraphics[width=0.85\textwidth]{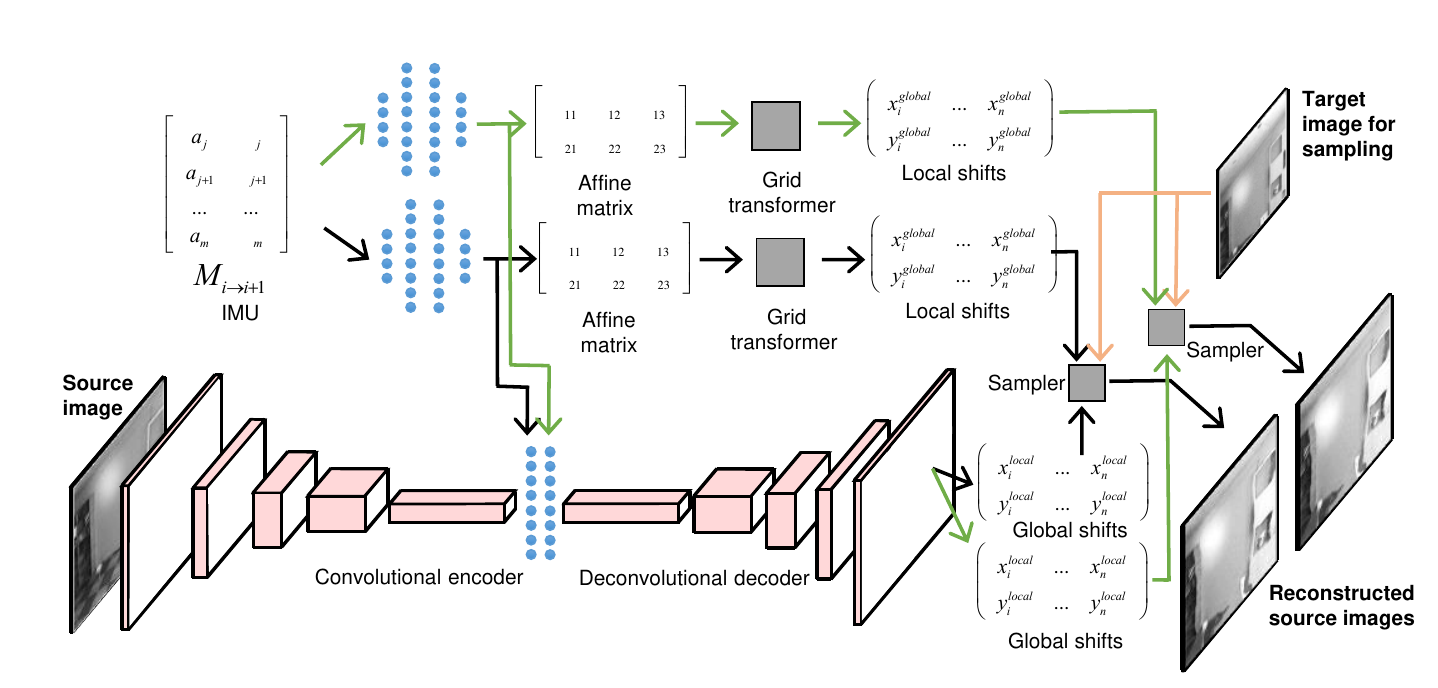}
	\caption{Network diagram with two hypotheses shown for brevity. We experimented with up to 8 hypothesis pathways.}
	\label{net_diagram}
\end{figure*}

\subsection{Visual Correspondence} \label{visual_correspondence}

The correspondence problem can be viewed as part of the more general problem of determining how images relate to one another. While the correspondence problem has been traditionally addressed through closed-form, analytical approaches (for example Deformable Pyramid Matching \cite{kim2013}; see \cite{scharstein2002} for a review of the correspondence problem), recent bio-inspired, deep neural network approaches that estimate the 3D spatial transformations between image pairs have begun to show increasing success.

Unsupervised, siamese-like deep network architectures such as those based on multiplicative interactions \cite{Memisevic2007, Ranzato2010, Memisevic2010, Hinton2011, Kivinen2011, Memisevic2013, wohlhart2015} and triplet learning rules \cite{wang2015} have been used successfully for relationship learning between RGB images at the cost of computational runtime. These architectures require expensive computations to be performed on both source and target images which can greatly increase model complexities due to the high-dimensional nature of image data.

Other learning approaches have relied on explicit supervised labeling such as the random decision forest based approaches of \cite{taylor2012,shotton2013, brachmann2014} and semantic segmentation approaches of \cite{long2015,hariharan2015}. These supervised approaches require expensive and time-consuming labeling that ultimately limits the size of useable datasets. The self-supervised visual descriptor approach of \cite{schmidt2017} used a learning rule that requires a priori labeling such that points in the source image (image at $t_{i}$) and target image (image at $t_{i+1}$) already be aligned and thus, for correspondence to have already been solved for the training set. Our approach is instead unsupervised and requires only raw image and IMU data (and intentional information, when available) for training.

\subsection{Optical Flow} \label{optical_flow}

Estimating optical flow is very similar to estimating correspondence. Usually, approaches to dense, large displacement optical flow require first estimating correspondences between images and then performing some type of variational joint energy-based optimization that includes assumptions for local smoothness \cite{horn1981}. Data-driven correspondence matching \cite{lowe2004, he2012} leave many outliers as the closest visual match between corresponding pixels in two images is not the same as the optical flow. In other words, correspondences for neighboring pixels can be quite noisy and show large discontinuities. SOA approaches such as EpicFlow \cite{revaud2015epicflow}, for example, apply such a methodology: first finding correspondences between images using \cite{Revaud2016}, and then computing dense optical flow. However, there are a number of other learning-based systems such as \cite{dosovitskiy2015, revaud2015epicflow, ilg2017 } that instead learn end-to-end models of optical flow from raw image inputs but still require extensive computation (see the last column of ÒTable \ref{Flow_results}Ò for runtimes). Other approaches such as \cite{pillai2017} take optical flow as an input for egomotion estimation (VIFlow could be used to generate this flow). 

\subsection{View Synthesis} \label{view_synthesis}
The view synthesis approach of \cite{zhou2016} learned local, pixel-level shifts in order to render new, unseen views of objects and scenes. That approach is similar to the mappings that are learned by the second pathway of our DE architectures. Besides different network structures and inclusion of a global spatial transformer module, the largest difference between their method and our own is that rather than learning to generate novel viewpoints of objects or scenes, we learn how to reconstruct a source image using pixel locations in a target image in order to compute estimates of correspondence and flow.

\subsection{Depth and Parallax} \label{depth_parallax}

While \cite{byravan2017} used depth data as input to their network, we do not provide VIFlow with depth. This is critical because given the 3D locations of points in the image scene, a 3D affine transformation can be directly performed to project points on the image plane at time $t_{i}$ to some time $t_{i+1}$. As VIFlow is designed for SWaP-constrained applications where only intensity information from a single imager might be available, our VIFlow network is input with only a single grayscale intensity image and uses its local pathway (see ÒSection \ref{approach}Ò and \cite{shamwell2017a,shamwell2017b} for more information) to infer depth and non-rigidity from a single 2D grayscale image (unlike the approaches of \cite{godard2017, garg2016, zhou2017, li2017, clark2017, finn2016, finn2016unsupervised} which all use color imagery which better enables explicit depth learning and \cite{byravan2017} which directly receives depth data and is additionally not trained based on sampled image reconstructive loss).

\subsection{Anamalous Independent Motion} \label{anomalous_motion}

For vision-aided egomotion estimation, independent motion sources in the visual scene introduce error to optical-flow derived estimates of egomotion. A number of approaches to independent motion detection take dense optical flow as an input \cite{Kitt2010, Ciliberto2012, Kumar2015, vasco2017}. The approach described in this paper could be used to estimate the flow that is fed into these independent motion detection algorithms or VIFlow could be augmented to directly separate ego-motion induced flow from flow induced by independently moving scene elements (see ÒSection \ref{section:anomaly_detection}Ò for more on VIFlow in this role).

\section{Approach} \label{approach}

Previous approaches have used ground-truth pose difference and ground-truth pose differences contaminated with heteroscedastic noise to show how an unsupervised network can learn to estimate a true 3D pose change with a global 2D affine transform plus localized, pixel-level coordinate shifts \cite{shamwell2017a, shamwell2017b}.

In contrast to \cite{shamwell2017a, shamwell2017b}, here we present results from experiments where networks were instead provided with raw sensor data in the form of a single grayscale image and IMU measurements. Additionally, we experimented with networks that also received intentional information in the form of a surrogate feed-forward motor signal (SFFMS) in place of an actual motor command as the benchmark datasets we used, KITTI \cite{Geiger2013} and EuRoC \cite{burris2016}, do not provide motor commands.

For IMU data, traditional fusion approaches assume a priori sensor and dynamics models with known intrinsic/extrinsic calibration parameters. Similarly, while intentional information may come from joysticked motor commands or command velocities freely available on a robotic system, traditional closed-form solutions to estimating the change in perception induced by commanded motion requires hand-crafted system models that relate motor inputs to changes in self-measured sensory perceptions. Contrastingly, our VIFlow network requires no explicitly defined forward model for motor commands affect sensory perceptions or IMU models and calibration parameters. Instead, VIFlow side-steps the need for hand-crafted modeling and learns how to combine heterogeneous inputs directly from the data.

VIFlow is designed to compute estimates for visuo-inertial flow and correspondence. 
It receives as input a single grayscale intensity image $I_{i}$ taken at time $t_{i}$ and an extra-visual estimate of camera motion $M_{i \rightarrow i+1}$ between time $t_{i}$ and time $t_{i+1}$. The goal for the network is to use the motion estimates $M_{i \rightarrow i+1}$ and the grayscale image $I_{i}$ to predict the new image coordinate in $I_{i+1}$ for each scene element captured in $I_{i}$. In other words, VIFlow learns to estimate the correspondence between pixels in images $I_{i}$ and $I_{i+1}$. 

The network architecture can be thought of as an extension of an autoencoder. However, rather than learning features by minimizing the reconstruction error between an input projected into feature-space and then re-projected into an output-space, VIFlow is trained by minimizing the reconstruction error between an input and and a reconstruction based on sampled values from a previously unseen target image $I_{i+1}$. 

Similar to the network in \cite{shamwell2017b}, VIFlow learns to generate several hypothesis reconstructions along a series of parallel pathways. These hypothetical reconstructions enable increased robustness to noisy inputs. To sample, VIFlow effectively is trained to draw from an unknown noise distribution using what we call a winner-take-all (WTA) Euclidean loss rule.

\subsection{Winner-Take-All (WTA) Loss Rule}

To utilize its the multiple hypothesis outputs, a WTA Euclidean loss rule (see  \cite{shamwell2017b} for more detail and justifications) is used to train VIFlow:

\begin{equation}
I_{r}^{*}(\theta,I_{t}) \longleftarrow \argmin_i \lVert I_{r}^{i}(\theta,I_{t}) - I_{s} \rVert ^ 2
\end{equation}
\begin{equation}
L(\theta,I_{t},I_{s}) = \lVert I_{r}^{*}(\theta,I_{t}) - I_{s} \rVert ^ 2 
\end{equation}

\noindent where $I_{r}^{*}$ is the lowest error hypothesis. Loss is then only computed for this one hypothesis and error is backpropagated only to parameters in that one pathway. Thus, only parameters that contributed to the winning hypothesis are updated and the remaining parameters are left untouched. While it may seem like such a loss rule may lead the network to only optimize a single pathway, in practice this was not the case and VIFlow networks continued to use multiple pathways throughout training and testing.

\subsection{Pathway 1: Global Shifter}
The first pathway of VIFlow (the top pathway shown in ÒFig. \ref{net_diagram}Ò) is the Global Shifter. Given a motion estimate (e.g. IMU data), it uses several fully-connected (FC) layers to approximate a 3D transformation as a 2D transformation by learning to compute the parameters for a 2D affine transformation matrix. The Global Shifter then applies this 2D affine transformation to generate expected coordinate shifts in the form of a HxWx2 grid that represents pixel locations at which to sample from in the target image (additional detail in ÒSection \ref{spatial_transformer}Ò below).

\subsection{Pathway 2: Local, Pixel-Level Shifter}

The second pathway of VIFlow (the bottom pathway shown in ÒFig. \ref{net_diagram}Ò) is the Local Shifter pathway. It receives a source image as input and uses a convolutional-deconvolutional encoder-decoder to also generate a HxWx2 output of pixel shifts. However, these shifts are intended to only modify the coordinate shifts calculated by the Global Shifter pathway for instances where the true motion of a scene element cannot be calculated using a single global 2D transform (e.g., for varied scene depths at pixels that differ from the dominant scene depth implicitly assumed by the 2D Global Shifter).

\subsection{Spatial Transformations} \label{spatial_transformer}

Spatial transformation in the form of a modified spatial transformer module \cite{Jaderberg2015} is an integral component of the VIFlow network architecture and an explanation helps to elucidate the workings of VIFlow. To perform a spatial transformation, we assume that output pixels are defined to lie on a regular grid $ G = \{G_{j}\} $ of pixels $ G_{j} = (x_{j}^t,y_{j}^t) $, forming an output feature map  $ V \in \Re^{H^{'} \times W^{'}} $ where H' and W' are the height and width of the grid. If we let $A(\theta,G_{j})$ represent a 2D affine transformation, then target coordinates are mapped to source coordinates according to 
  
\begin{equation}
\left(  \begin{array}{ccc}
x_{j}^s \\
y_{j}^s  \end{array} \right) =   A(\theta,G_{j}) = \left[  \begin{array}{ccc}
\theta_{11} & \theta_{12} & \theta_{13} \\
\theta_{21} & \theta_{22} & \theta_{23}  \end{array} \right] 
\left(  \begin{array}{ccc}
x_{j}^t \\
y_{j}^t  \\
1 \end{array} \right) \\
\end{equation}

\noindent where $\left(x_{i}^t, y_{i}^t\right)$ are target coordinates in the output feature map, $(x_{i}^s, y_{i}^s)$ are the source coordinates in the input feature map, and $\theta$ is the 2D affine transformation matrix.

VIFlow's Global Shifter pathway learns how to compute the matrix $\theta$ for a motion estimate $M_{i \rightarrow i+1}$ and then applies the transform to generate global coordinate estimates $[X_{global}, Y_{global}, 1]^{T}$.

If all pixels are of the same depth, then the Global Shifter pathway would be able to accurately project pixels to their correct positions post-movement. However, because this is not the case, the best result the Global Shifter can accomplish is the correct projection of points that belong to some dominant plane. The Local Shifter is able to apply corrections to the coordinate changes computed by the Global Shifter to allow for differing object depths in a scene and non-rigidity.

Given a source image $I_{i}$, VIFlow's Local Shifter pathway learns to compute $[X_{local}, Y_{local}, 1]^{T}$ which are localized shifts to be summed with the global shifts $[X_{global}, Y_{global}, 1]^{T}$ computed by the Global Pathway. Thus, VIFlow generates its final coordinate locations as

\begin{equation}
[X_{shift}, Y_{shift}, 1]^{T} =  [X_{global}+X_{local}, Y_{global}+Y_{local}, 1]^{T}
\end{equation}

\noindent and then performs bilinear sampling to produce a reconstruction image $Ir$ by sampling an unseen image $I_{i+1}$ at coordinates $(x,y) \in [X_{shift}, Y_{shift}]$.

\section{Methods} \label{methods}

\begin{figure} 
  \centering    
  \begin{subfigure}[EuRoC]{0.23\textwidth}  
    \label{euroc_kmeans}
 	\def\svgwidth{40mm}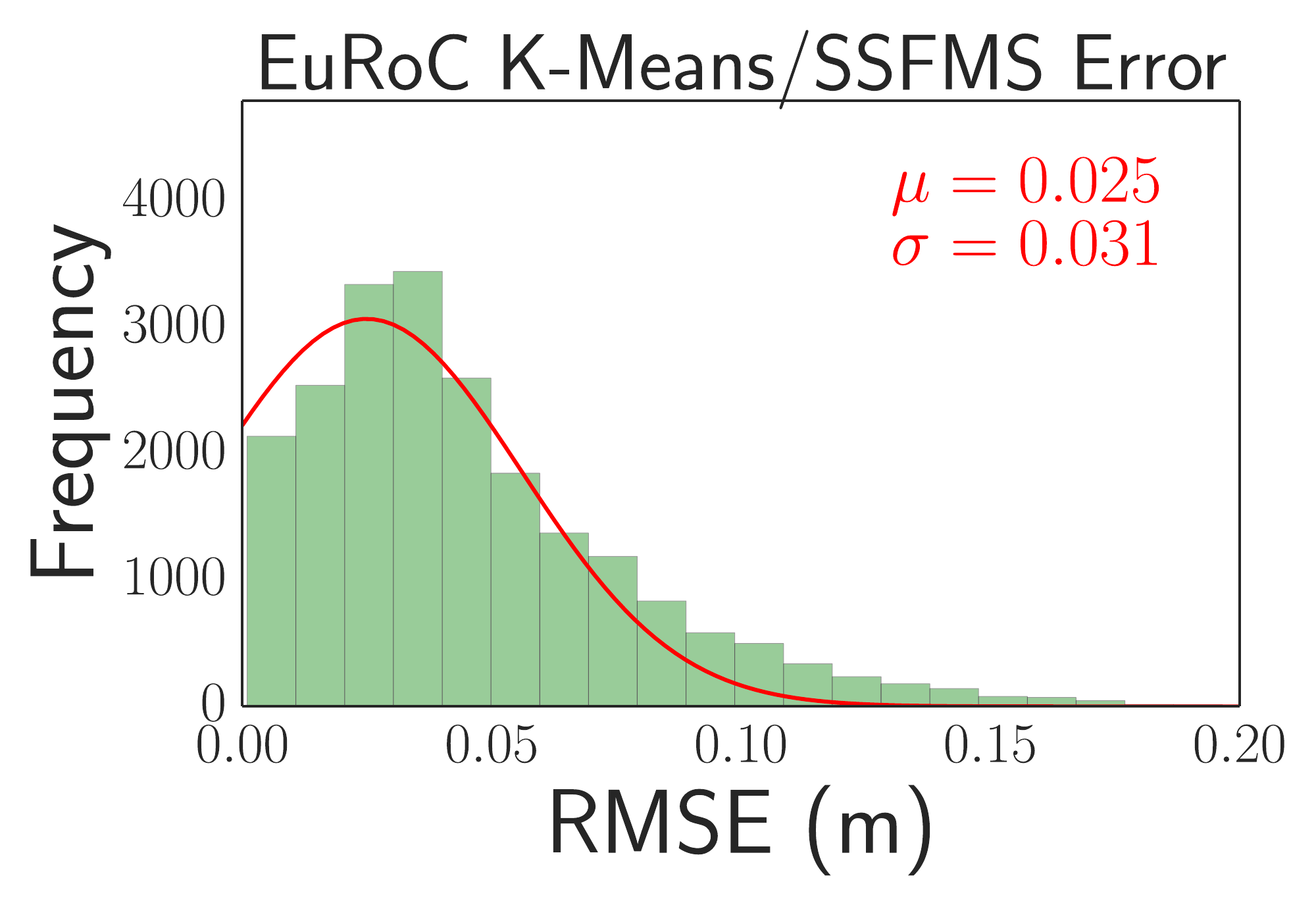
 	\caption{EuRoC}
  \end{subfigure}	
  \begin{subfigure}[KITTI]{0.23\textwidth} 
    \label{kitti_kmeans}
	\def\svgwidth{40mm}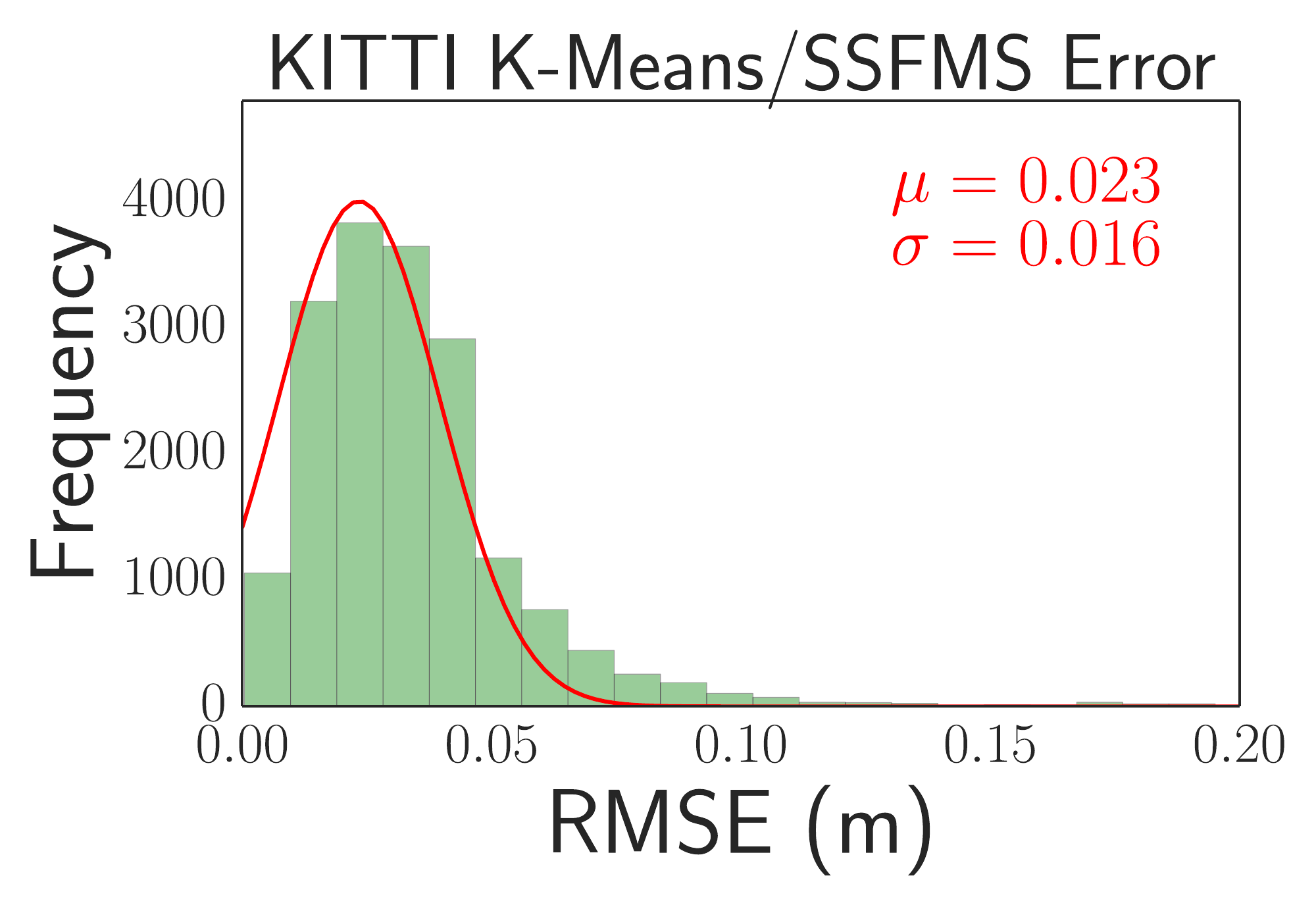	
    \caption{KITTI}
  \end{subfigure}		
  \caption{Histogram of error between the ground truth and K-Means cluster to which that exemplar was assigned. Note that error in EuRoC is heavier tailed compared to error in KITTI, indicative of larger variance in the motion transforms in EuRoC. }
  \label{hist_kmeans}
\end{figure}

We designed networks with motion information taken from the following sources:

\begin{enumerate}

\item \textbf{Raw Measurements from an Inertial Measurement Unit (IMU)}: The IMU data recorded between capture times for the source and target images were input to the network as the extra-visual motion transform.
\item \textbf{Surrogate Feed-forward Motor Signals (SFFMS)}: Because direct motor-command inputs were not available, a K-Means fitting of clusters was used on the ground truth position differences to generate noisy estimates of the approximate direction of motion (see ÒFig. \ref{hist_kmeans}Ò for the associated error). We evaluated networks with these inputs and present results to show how a higher-level signal that encodes intentional information (even if it is noisy) can affect network performance.

\item \textbf{IMU+SFFMS}: IMU and SFFMS data as above were both input.
\end{enumerate}

\subsection{Datasets and Data Generation}

Because we require IMU data as well as ground truth pixel end points (or a means to calculate ground truth; see ÒAppendix \ref{appendix_euroc_gt}Ò), we were limited in our choice of public datasets. For example, we could not use the Middlebury and MPI-Sintel visual flow benchmark datasets as they do not include IMU data. 

We thus used the KITTI \cite{Geiger2013} and EuRoC \cite{burris2016} datasets for evaluation. For each image in KITTI and EuRoC, we cropped the middle 224x224 pixel region for inputs to the network. Details on each dataset and dataset specific data preparation follow below.

\subsubsection{EuRoC MAV}

EuRoC contains data collected with a VI-Sensor \cite{nikolic2014} which captures stereo image data at 20 Hz and IMU data at 200 Hz. 

The total number of usable exemplars \textit{Vicon Room} scenarios was $12,781$ which is significantly smaller than the KITTI odometry dataset. We thus elected to augment the V01 datasets in EuRoC by including not only pairs of sequential frames, but pairs separated by up to four frames. This resulted in $26,976$ total examples, of which $80\%$ $(21,588)$ were used for training and $20\%$ $(5,388)$ were used for testing.

For the IMU and IMU+SFFMS conditions, because the lookahead could be anywhere from one to four frames and thus anywhere between 50ms to 200ms, the IMU inputs for the EuroC models used a vector of size 50x6 where for all exemplars regardless of lookahead size, the first 10 entries correspond to the 50 ms prior to the image capture and the next 10 entries correspond to the 50ms following capture. For exemplars that were only one ahead, the remainder of the vector were zeros. For lookaheads of 2 the last 20 entries were zeros; for lookaheads of three the last 10 were zeros; and finally for lookaheads of four the vector was fully populated.

To generate the surrogate SFFMS, we performed K-Means clustering on the ground truth position differences to generate 20 clusters which were encoded as a one-hot vector.

\subsubsection{KITTI Odometry}

For KITTI, we used sequences $00-10$ excluding sequence $03$ because the corresponding raw file $2011\_09\_26\_drive\_0067$ was not online at the time of publication. This resulted in a total of $20,976$ image pairs for training (pairs of images for which corresponding IMU data was unavailable or had inconsistent timestamps were excluded). In all experiments, we randomly selected an image for the source and used the successive image for the target. Corresponding 100Hz IMU data was collected from the KITTI raw datasets and the preceding 100 ms and following 100 ms of IMU data was included for each example yielding a length 20x6 vector. SSFMS were generated as above.

\subsection{Network Parameters and Training Procedures}

For the VIFlow-IMU networks, four FC layers of size $512$, $4096$, $4096$, and $512$ were used to generate the 2x3 affine transformation matrices. For the IMU+FFMS configurations which  had two sources of extra-visual motion estimates as described above, each extra-visual modality was processed through four FC layers of $512$, $4096$, $4096$, and $512$ before being concatenated into a vector of length $1024$.

The convolutional-deconvolutional encoder-decoder that composed the Local Shifter pathway used 5x5 convolutional kernels with a stride of two. The encoder used five layers of $32$, $64$, $128$, $256$, and $512$ filters and the decoder was reversed, using $512$, $256$, $128$, $64$, and $32$ filters. All results described in this paper used a Local Shifter pathway with these parameters.

As shown in ÒFig. \ref{net_diagram}Ò, the output of fifth convolutional layer is concatenated with the last FC layer of the Global Shifter pathway and was then fed into a single FC layer of size $4096$ before being fed into the first deconvolutional decoder layer.

We trained three networks for each condition and dataset and all results presented are from the highest performing network for each condition. Networks were trained on a desktop computer with a 3.00 GHz Intel i7-6950X processor and Nvidia Titan X GPUs. 
  
\section{Evaluation} \label{evaluation}

Predicted pixel correspondence between source and target images was evaluated against ground truth correspondence, correspondence computed by the DeepMatching algorithm \cite{Revaud2016}, and correspondence computed by the Deformable Spatial Pyramid Matching algorithm (DSP) \cite{kim2013} on the EuRoC MAV \cite{burris2016} and KITTI SceneFlow 2015 datasets \cite{Geiger2013}.

While VIFlow is most similar to a dense correspondence network that estimates the nearest matches between pixels in two images and does not explicitly use regularization on computed matches, we also evaluated VIFlow against several deep optical flow networks that are designed to perform additional regularization and outlier removal: EpicFlow \cite{revaud2015epicflow}, FlownetC \cite{dosovitskiy2015}, and Flownet2 \cite{ilg2017}.

It should be emphasized that the approaches we use to benchmark the VIFlow networks do not receive extra-visual motion inputs as we were unable to identify an approach in the literature with our exact input/output domain.

\subsection{Ground Truth}

For KITTI, ground-truth optical flow is provided for a subset of images (SceneFlow 2015) and we used exemplars from the training set to test our network (of the 200 training images, only 140 mapped to raw data from which we could extract IMU data and had IMU data without temporal discontinuities).

For EuRoC, pixel-level ground-truth is not available and needed to be calculated as described in ÒAppendix \ref{appendix_euroc_gt}Ò.

\def\arraystretch{1.25}
\begin{table*}[t]
\caption{Benchmark results for our MHIDE networks compared to FlownetC  \cite{dosovitskiy2015}, Flownet2 \cite{ilg2017}, EpicFlow \cite{revaud2015epicflow}, DeepMatching \cite{Revaud2016}, and DSP \cite{kim2013} on the KITTI \cite{Geiger2013} and EuRoC MAV \cite{burris2016} datasets for endpoint error (EPE). The last columns are calculated runtime/performance quotients for KITTI and EuRoC which are the ratios of AEPE to inverse runtime scaled by $0.01$. }
\label{Flow_results}
\begin{center}
\begin{tabular}{l*{13}{c}}
\hline
 & & & \multicolumn{3}{c}{\textit{KITTI (EPE)}} & \multicolumn{3}{c}{\textit{EuRoC (EPE)}} & \multicolumn{1}{c}{\textit{Runtime}} & \multicolumn{2}{c}{\textit{Perf./Run. Quotient}} \\
\cmidrule(lr){4-6} 
\cmidrule(lr){7-9} 
\cmidrule(lr){10-10}
\cmidrule(lr){11-12}

Algorithm & Hypoths. & Dense & $\mu$ & $\sigma$ & Med. & $\mu$ & $\sigma$ & Med.  & Ms & KITTI & EuRoC\\
\hline
\hline
FlownetC \cite{dosovitskiy2015} & N/A & Y & 4.00 & 3.41 & 3.28 & 4.75 & 6.92 &	2.78 & 88.7 & 3.55 & 4.21 \\
Flownet2 \cite{ilg2017} & N/A & Y & 0.52 & 0.71 & 0.34 & 0.6 & 0.73 & 0.2 & 197.2 & 1.03 &	1.18 \\
EpicFlow \cite{revaud2015epicflow} & N/A & Y & 1.86 & 4.05 & 0.88  & 0.55 & 0.72 & 0.15 & 2845.2 & 52.99	& 15.65 \\
DeepMatching  \cite{Revaud2016} \footnotemark  & N/A & N & 1.76 & 2.09 & 1.40 & 3.87 & 6.83 & 2.09 & 411.5 & 28.84	& 15.93 \\
DSP \cite{kim2013} & N/A & Y & 5.11 & 4.76 & 3.91 & 4.37 & 8.39 & 1.94 & 1252 & 95.3	 & 54.71 \\
Identity & N/A & N/A & 8.46 & 7.51 & 7.52 & 20.63 & 22.00 & 13.49 & N/A & N/A & N/A \\
\hline
\hline
VIFlow-IMU & 1 & Y & 5.02 & 6.66 & 3.05 & 5.98 & 5.61 & 4.39 & 4.7 & 0.24	& 0.28\\
VIFlow-IMU & 4 & Y & 4.80 &	6.53	 & 3.11 & 4.71 & 4.53 & 3.38 & 6 & 0.29 & 0.28\\
VIFlow-IMU & 8 & Y & 4.54 & 6.51 & 2.86 & 3.87 & 3.77 & 2.83 & 8.2 & 0.37	& 0.32 \\
\hline
VIFlow-SFFMS & 1 & Y & 6.75 & 8.10 & 4.14 & 12.78 & 11.77 & 9.71 & 4.7 & 0.32	& 0.6 \\
VIFlow-SFFMS & 4 & Y & 4.20 & 5.82 & 2.56 & 8.12 & 7.22 & 5.94 & 6.1 & 0.26 & 0.5 \\
VIFlow-SFFMS & 8 & Y & 3.47 &  5.49 & 2.08 & 6.19 & 5.58 & 4.67 & 8.1 & 0.28 & 0.5\\
\hline
VIFlow-IMU + SSFMS & 1 & Y & 3.28 & 6.15 & 1.61 & 5.23 & 4.83 & 3.95 & 5 & 0.16 & 0.26 \\
VIFlow-IMU + SSFMS & 4 & Y & 2.98 & 6.11 & 1.33 & 3.82 & 3.71 & 2.82 & 6.4 & 0.19 & 0.24\\
VIFlow-IMU + SSFMS & 8 & Y & 2.91 & 6.14 & 1.24 & 3.29 &	3.37 & 2.38 & 8.5 & 0.25	 & 0.28\\

\end{tabular}
\end{center}
\end{table*}

\begin{figure}[th]
	\begin{minipage}{.49\textwidth}
		\begin{subfigure}{\linewidth}
    			\begin{subfigure}{0.24\linewidth}
  	   			\includegraphics[width=\linewidth]{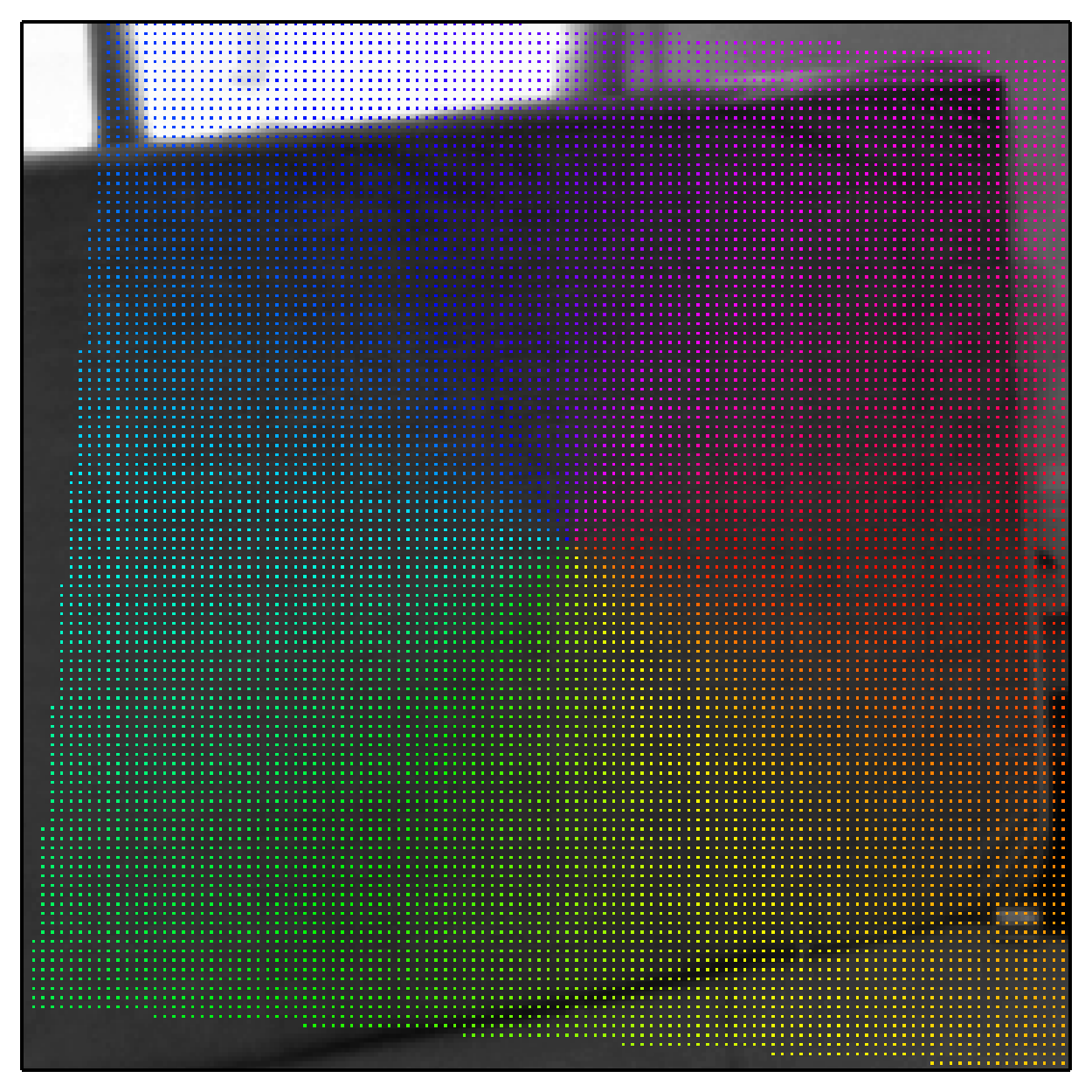}
  	 		\end{subfigure}%
    			\begin{subfigure}{0.24\linewidth}
  	   			\includegraphics[width=\linewidth]{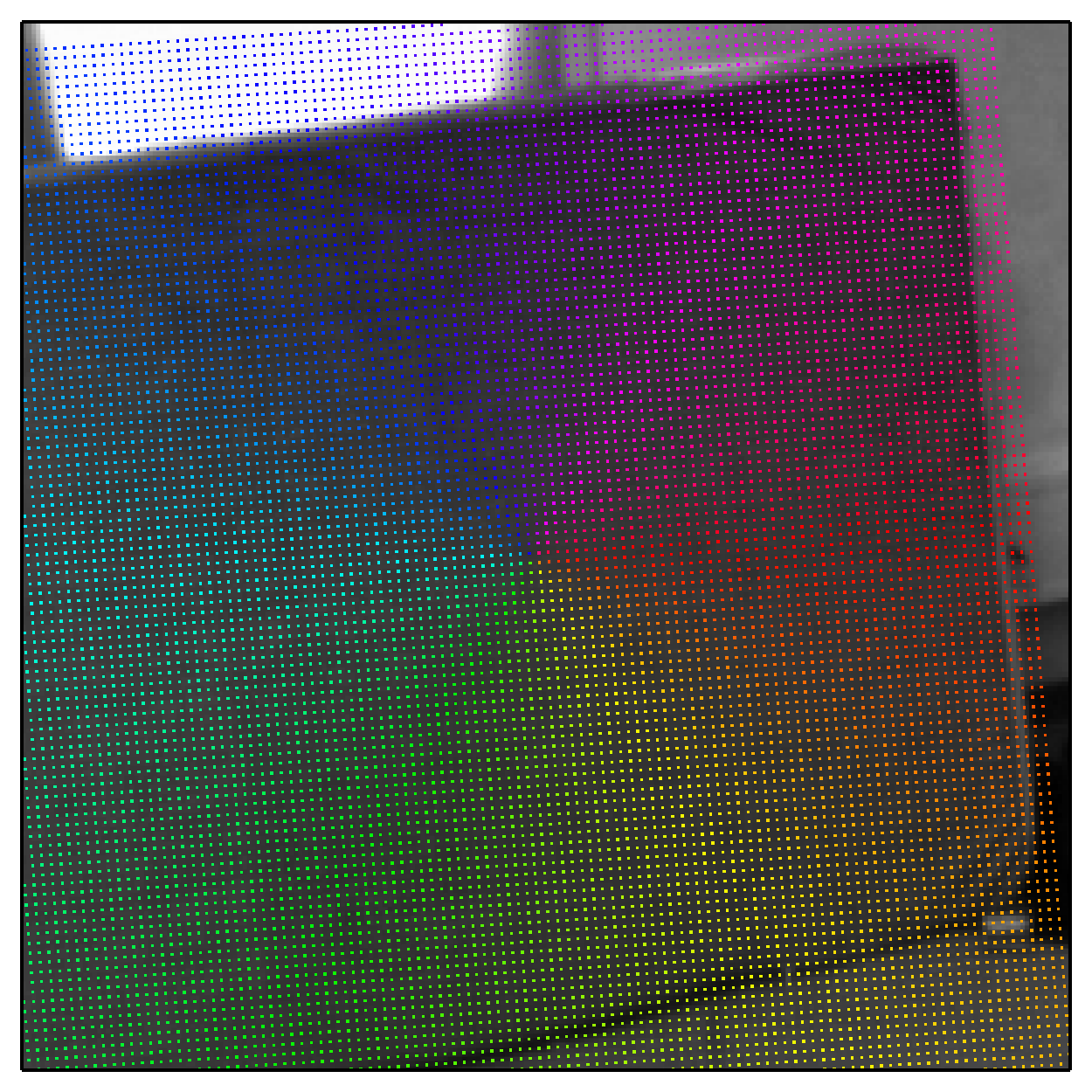}
  	 		\end{subfigure} \hfill
    			\begin{subfigure}{0.24\linewidth}
  	   			\includegraphics[width=\linewidth]{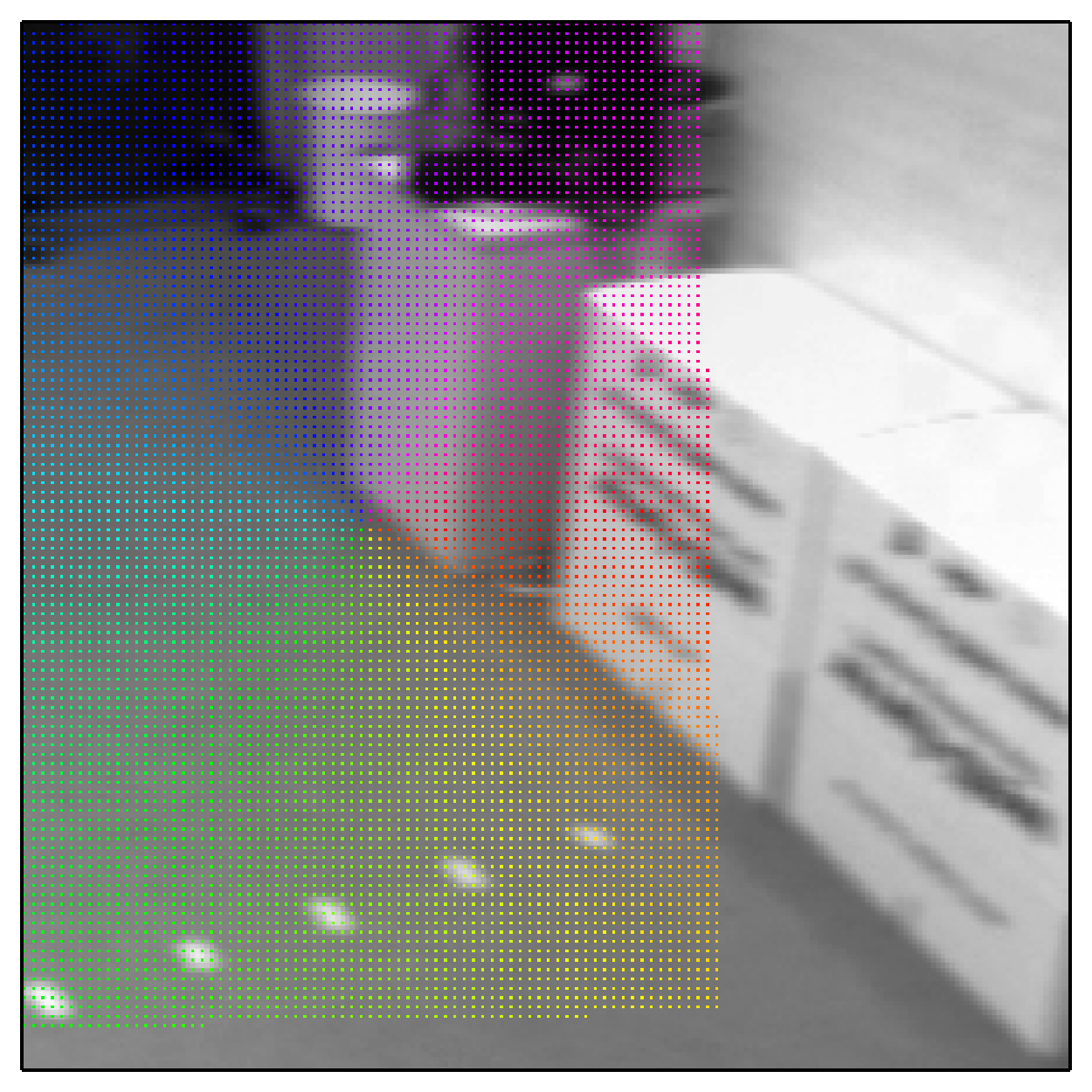}
  	 		\end{subfigure}%
    			\begin{subfigure}{0.24\linewidth}
  	   			\includegraphics[width=\linewidth]{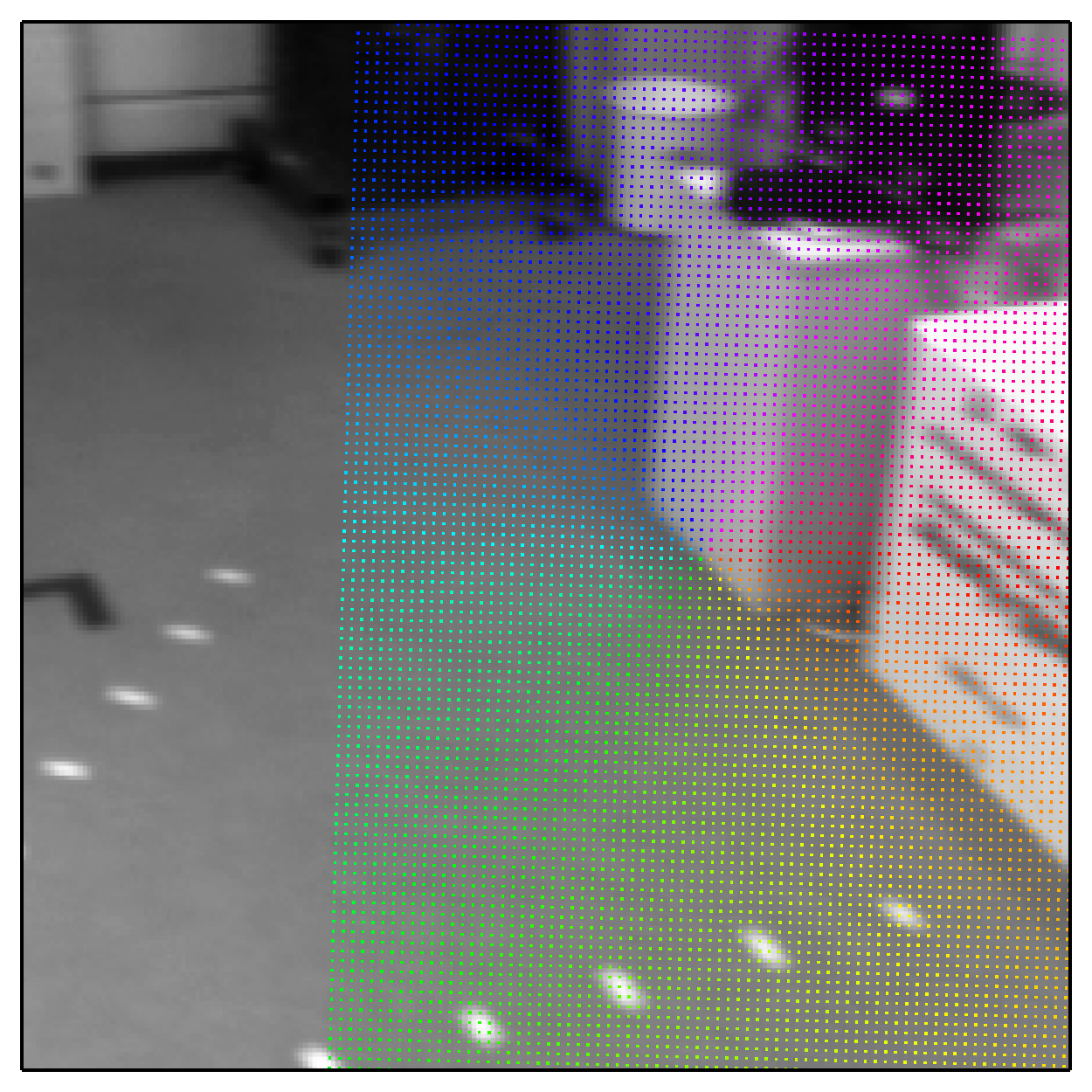}
  	 		\end{subfigure}%
	 	 \end{subfigure}%
	 \end{minipage}
	\begin{minipage}{.49\textwidth}
		\begin{subfigure}{\linewidth}
    			\begin{subfigure}{0.24\linewidth}
  	   			\includegraphics[width=\linewidth]{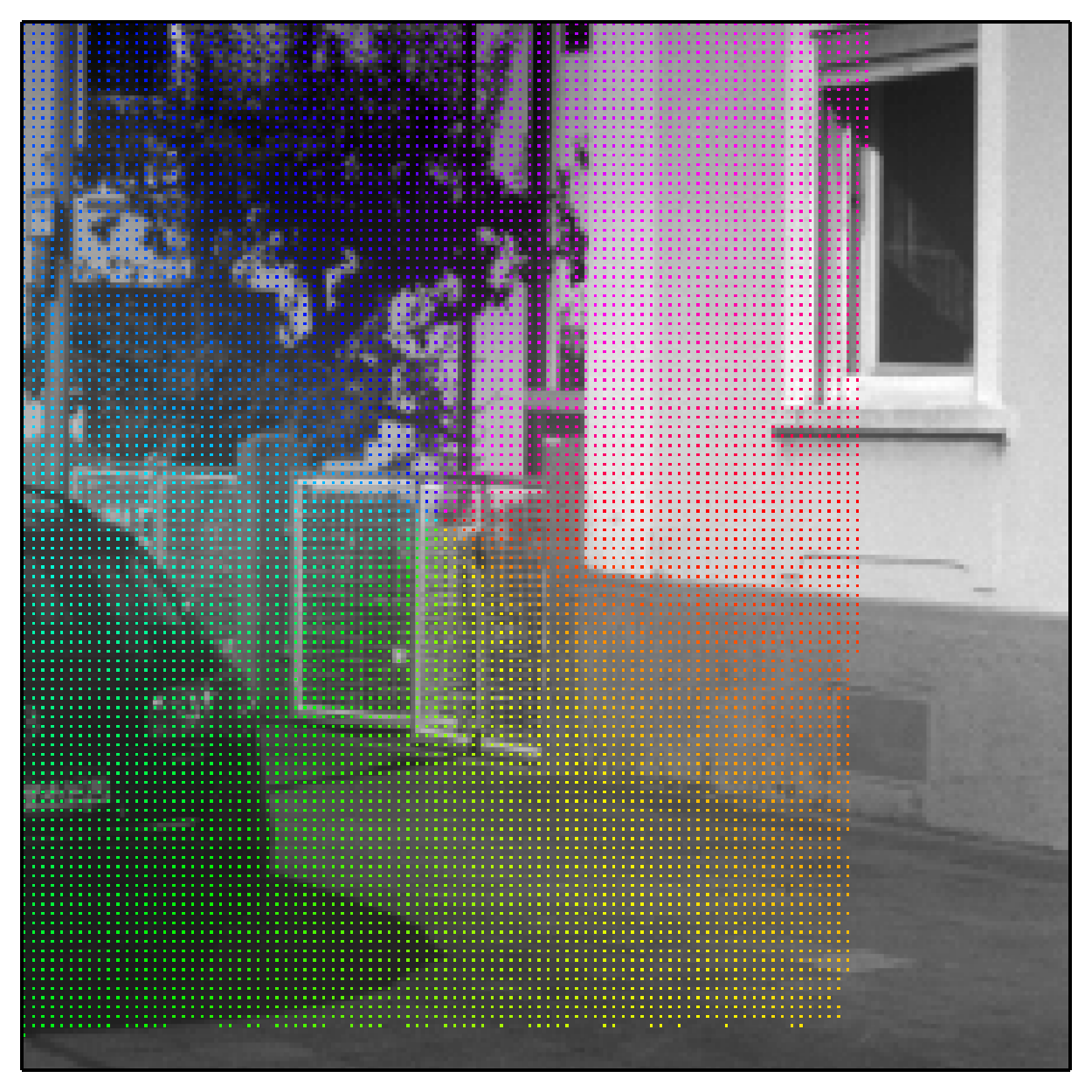}
  	 		\end{subfigure}%
    			\begin{subfigure}{0.24\linewidth}
  	   			\includegraphics[width=\linewidth]{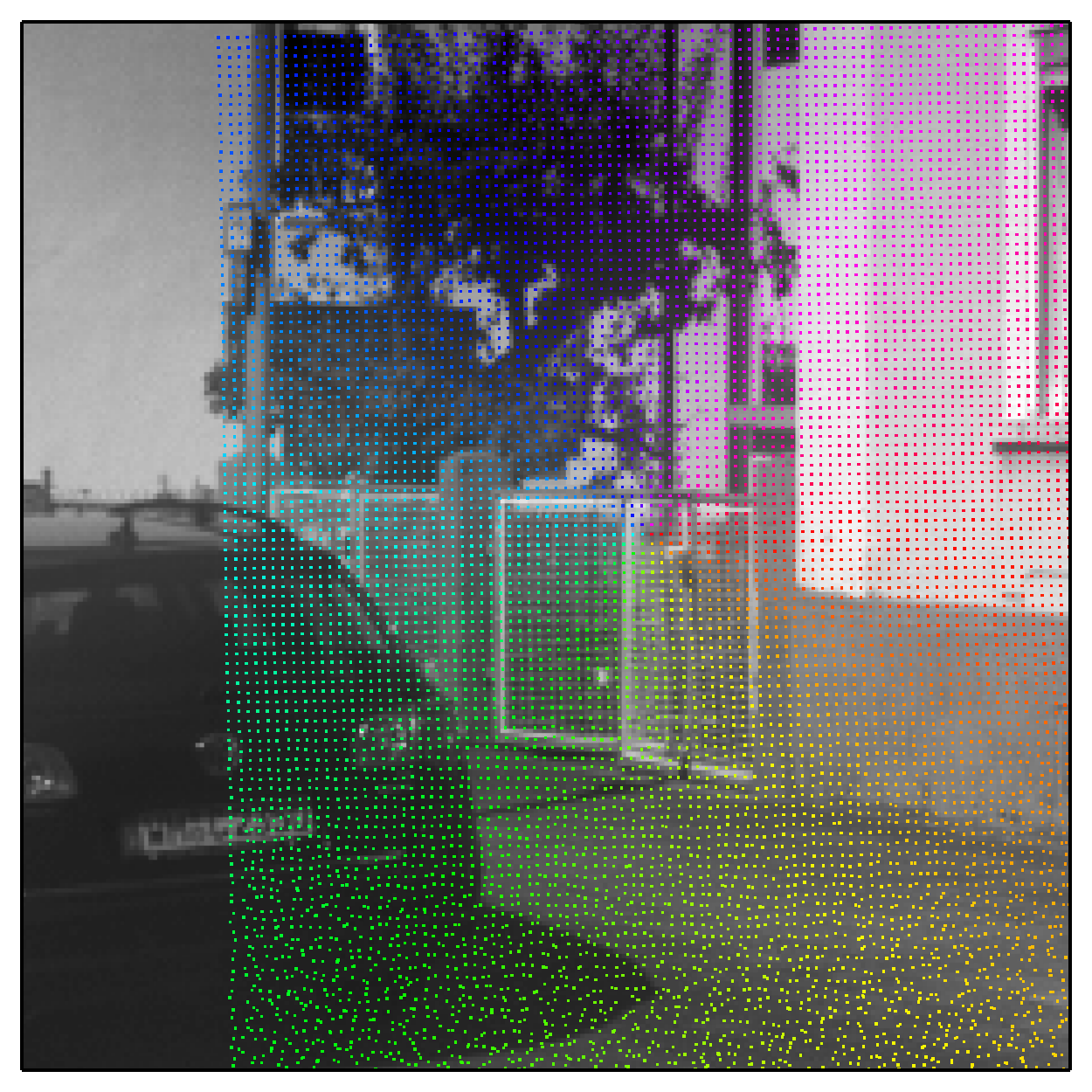}
  	 		\end{subfigure} \hfill
    			\begin{subfigure}{0.24\linewidth}
  	   			\includegraphics[width=\linewidth]{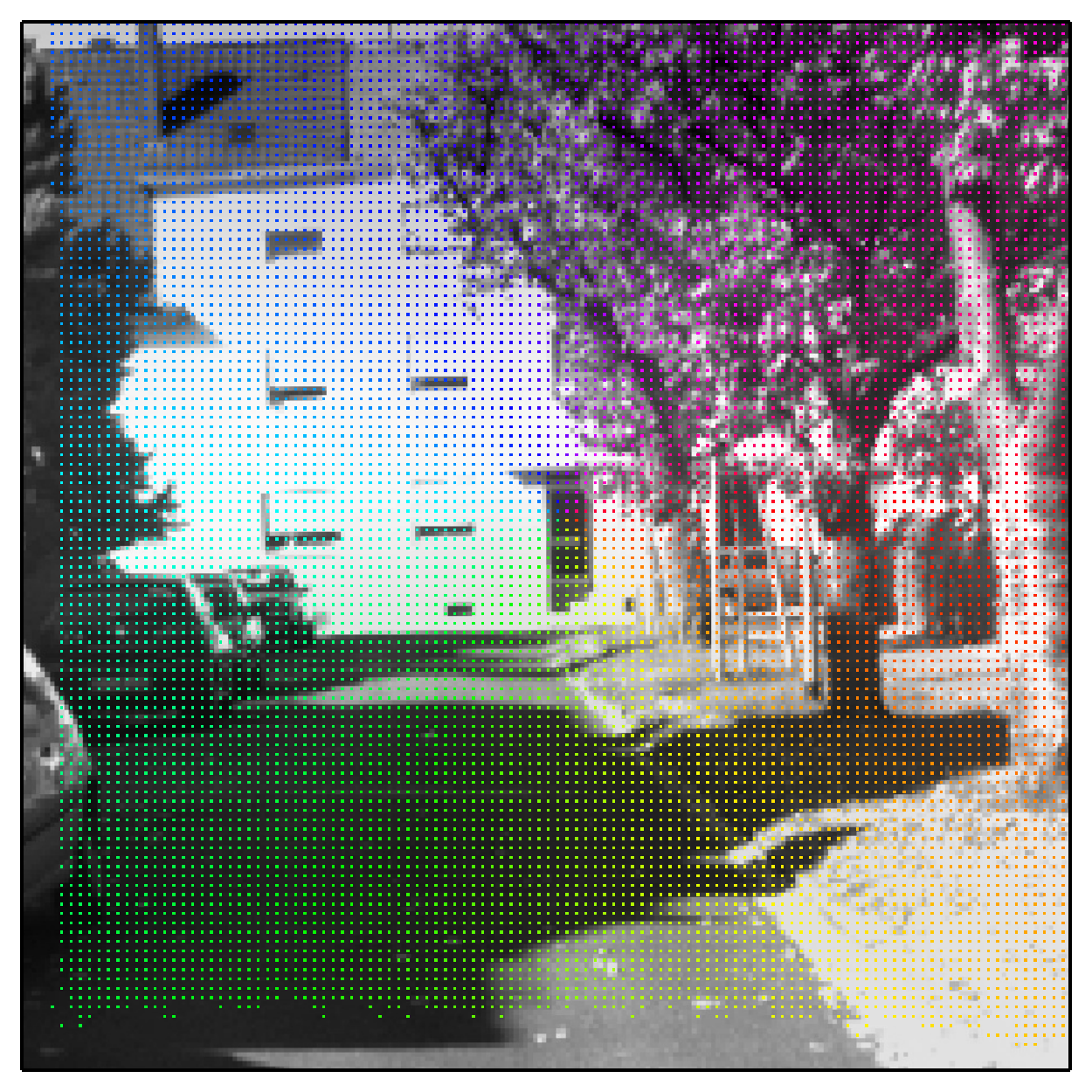}
  	 		\end{subfigure}%
    			\begin{subfigure}{0.24\linewidth}
  	   			\includegraphics[width=\linewidth]{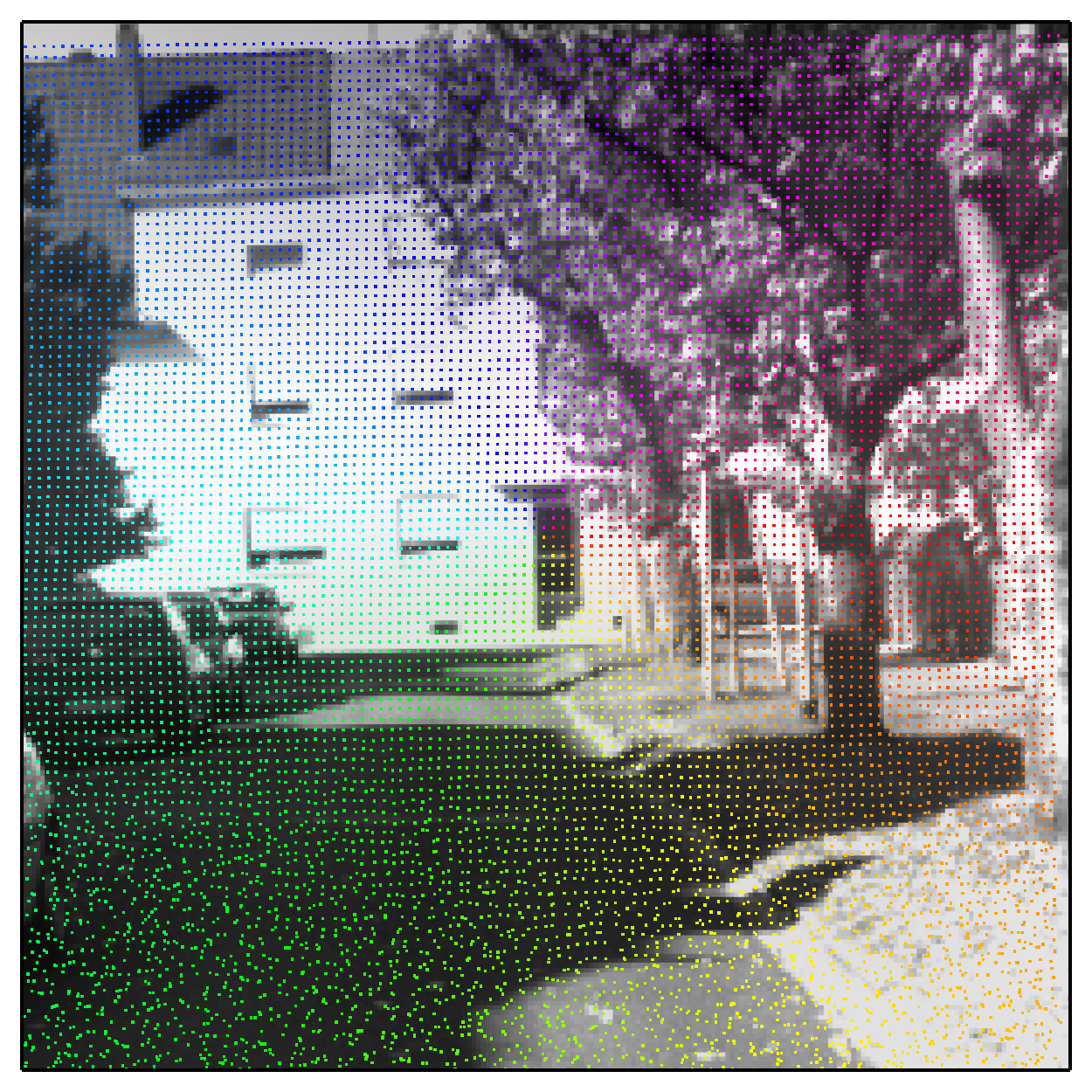}
  	 		\end{subfigure}%
	 	 \end{subfigure}%
	 \end{minipage}	 
	\caption{Sample EuRoc (top row) and KITTI (bottom row) correspondence results. Note that only every only keypoint (horizontally and vertically) is shown and the actual, unaltered output is fully dense. }		
	\label{matching_results}
\end{figure}  

\section{Results and Discussion} \label{results}

\subsection{Correspondence and Flow}

ÒTable \ref{Flow_results}Ò presents our results from the various VIFlow networks on the KITTI SceneFlow 2015 \cite{Geiger2013} and EuRoC MAV \cite{burris2016} datasets compared to FlownetC \cite{dosovitskiy2015}, Flownet2 \cite{ilg2017}, EpicFlow \cite{revaud2015epicflow}, DeepMatching \cite{Revaud2016}, DSP \cite{kim2013}, and an identity mapping. The identity mapping results were computed by assuming a flow of zero and are shown only to provide perspective on the quantitative results presented in ÒTable \ref{Flow_results}Ò.

The VIFlow networks were the fastest of all the approaches tested. It should of course be noted that in comparison to the other approaches, the VIFlow networks were the only approaches provided with a motion prior in the form of IMU data, a SFFMS, or both. We calculate a performance/runtime quotient, which are the ratios of average end-point error (AEPE) to inverse runtime scaled by $0.01$, and the VIFlow networks generate the best quotients (lower is better) indicating high efficiency. However, VIFlow was significantly outperformed in AEPE by the computationally heavier EpicFlow and Flownet2 networks in all conditions.

\subsubsection{VIFlow-IMU}

The $8$ hypothesis VIFlow-IMU architecture outperformed DSP on KITTI and EuRoC, approximately equaled the performance of DM on EuRoC, and was outperformed by DM on KITTI.

\subsubsection{VIFlow-SSFMS}
Even in the $8$ hypothesis case, VIFlow-SFFMS was outperformed by the correspondence approaches on EuRoC but managed the opposite on KITTI. In general, VIFlow-SFFMS networks showed the greatest performance improvements as a function of hypothesis pathways and event eventually surpassed VIFlow-IMU for KITTI in the $8$-hypothesis case. These reults are likely related to the heavier tailed error distributions for EuRoC seen in ÒFig. \ref{hist_kmeans}Ò compared to KITTI ($\sigma=0.031$ vs. $\sigma=0.016$) for SSFMS inputs which suggests that additional motion variability in EuRoC led to increased noise for the K-Means estimates and the network was not able to sufficiently learn this distribution with the number of hypotheses allotted.

The performance difference between KITTI and EuRoC suggest that their performance is tied to the variability in the underlying motion being experienced by the respective vehicles, with the lower-variability KITTI motions allowing for a better fitting K-Means clustering and the opposite for EuRoC.

\subsubsection{VIFlow-IMU+SSFMS}
For KITTI, the single hypothesis version of VIFlow-IMU+SSFMS showed a significant performance increase compared to both VIFlow-IMU and VIFlow-SSFMS. The combined performance increase of the IMU+SSFMS networks suggest that the information in the IMU measurements and SSFMS signals convey complementary information that enable a joint-reduction in uncertainty. However, it should again be noted that our SSFMS inputs were not actual motor signals (see ÒSection \ref{motor_limitations}Ò for further discussion on this limitation).

\begin{figure}[th]
	\begin{center}
	\begin{minipage}{.3\textwidth}
		\begin{subfigure}{0.33\linewidth}
			\begin{subfigure}{\linewidth}
				\includegraphics[width=\textwidth]{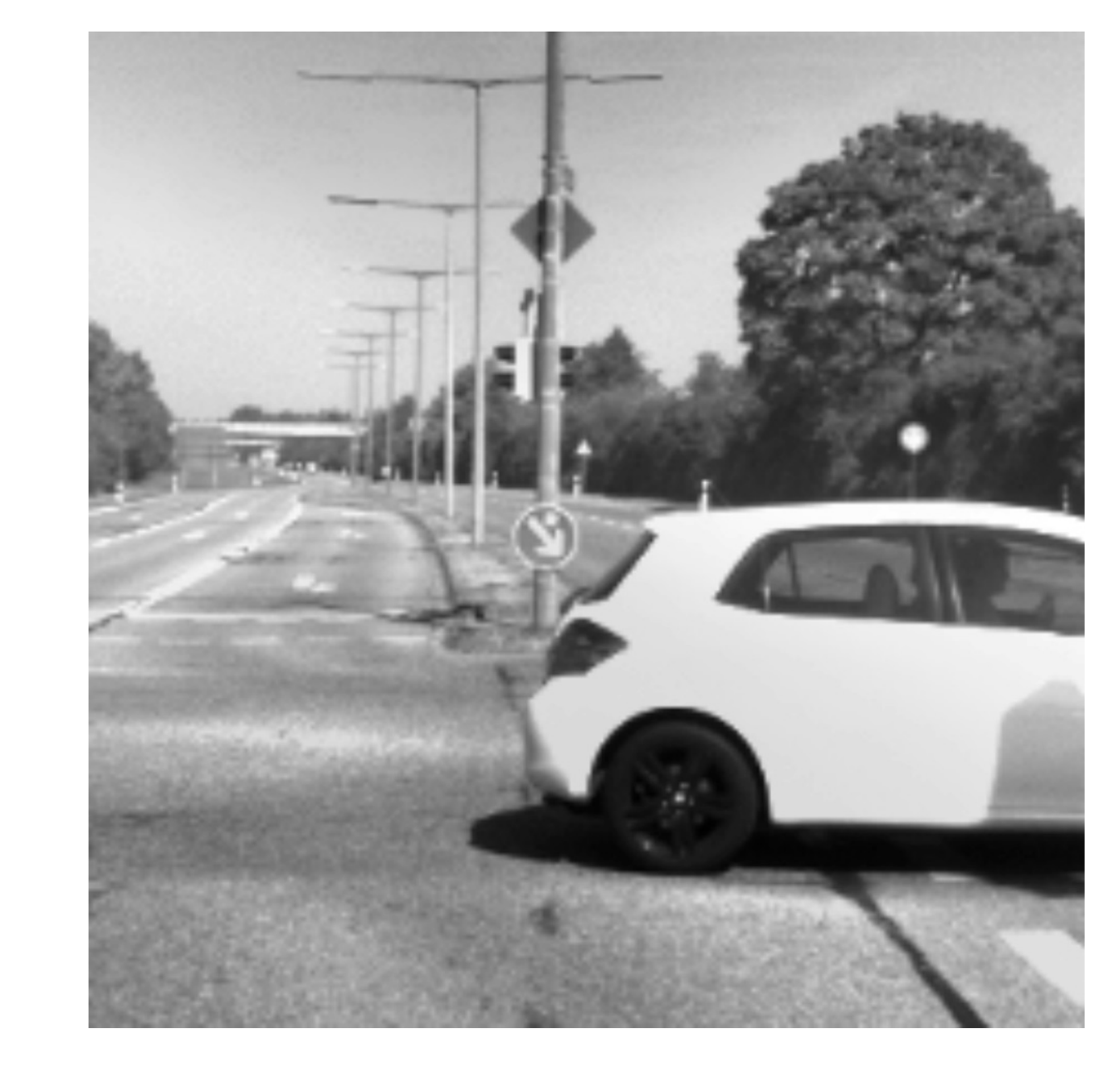}
	  		\end{subfigure}%
			\begin{subfigure}{\linewidth}
				\includegraphics[width=\textwidth]{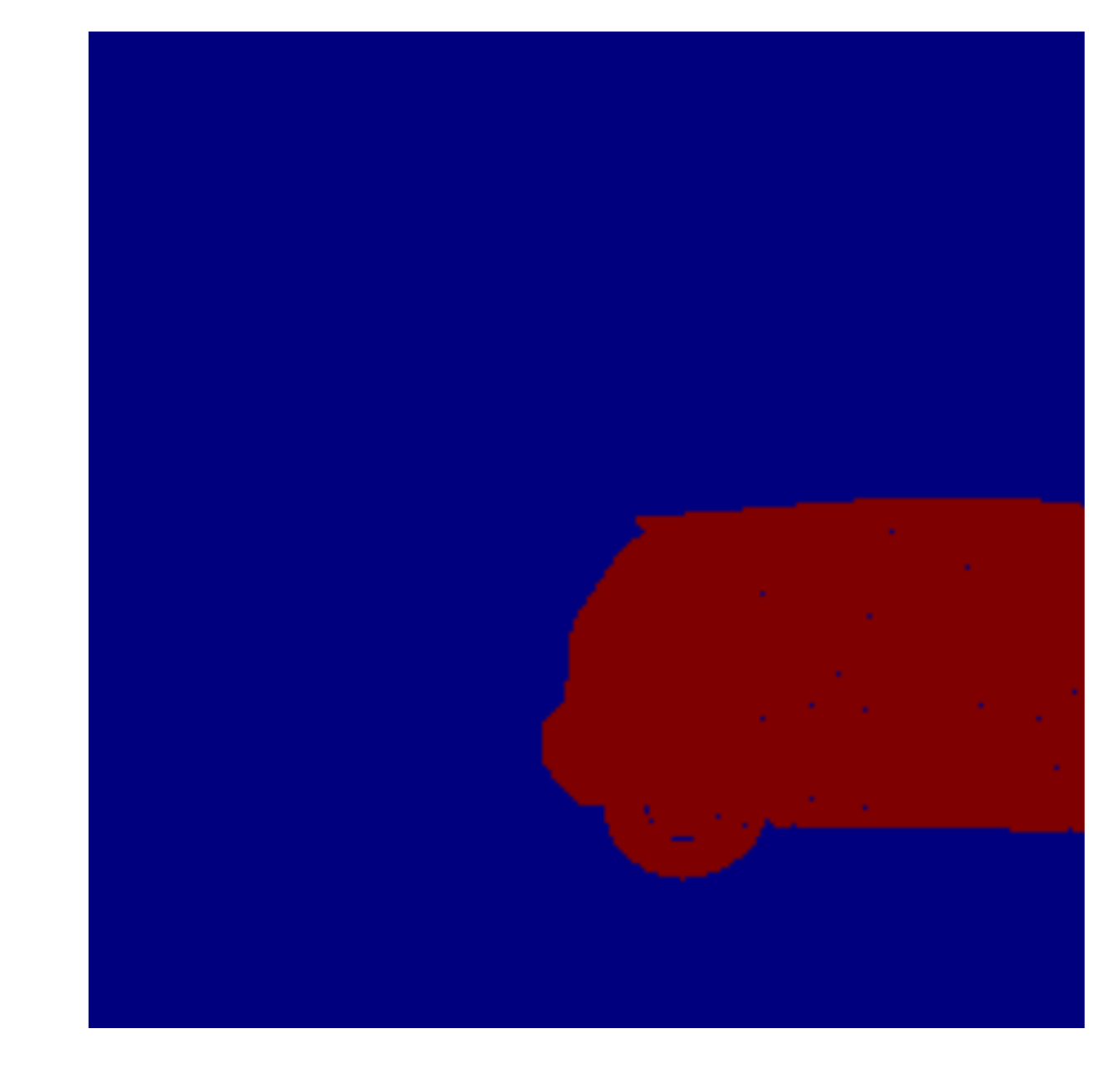}
	  		\end{subfigure}%
			\begin{subfigure}{\linewidth}
				\includegraphics[width=\textwidth]{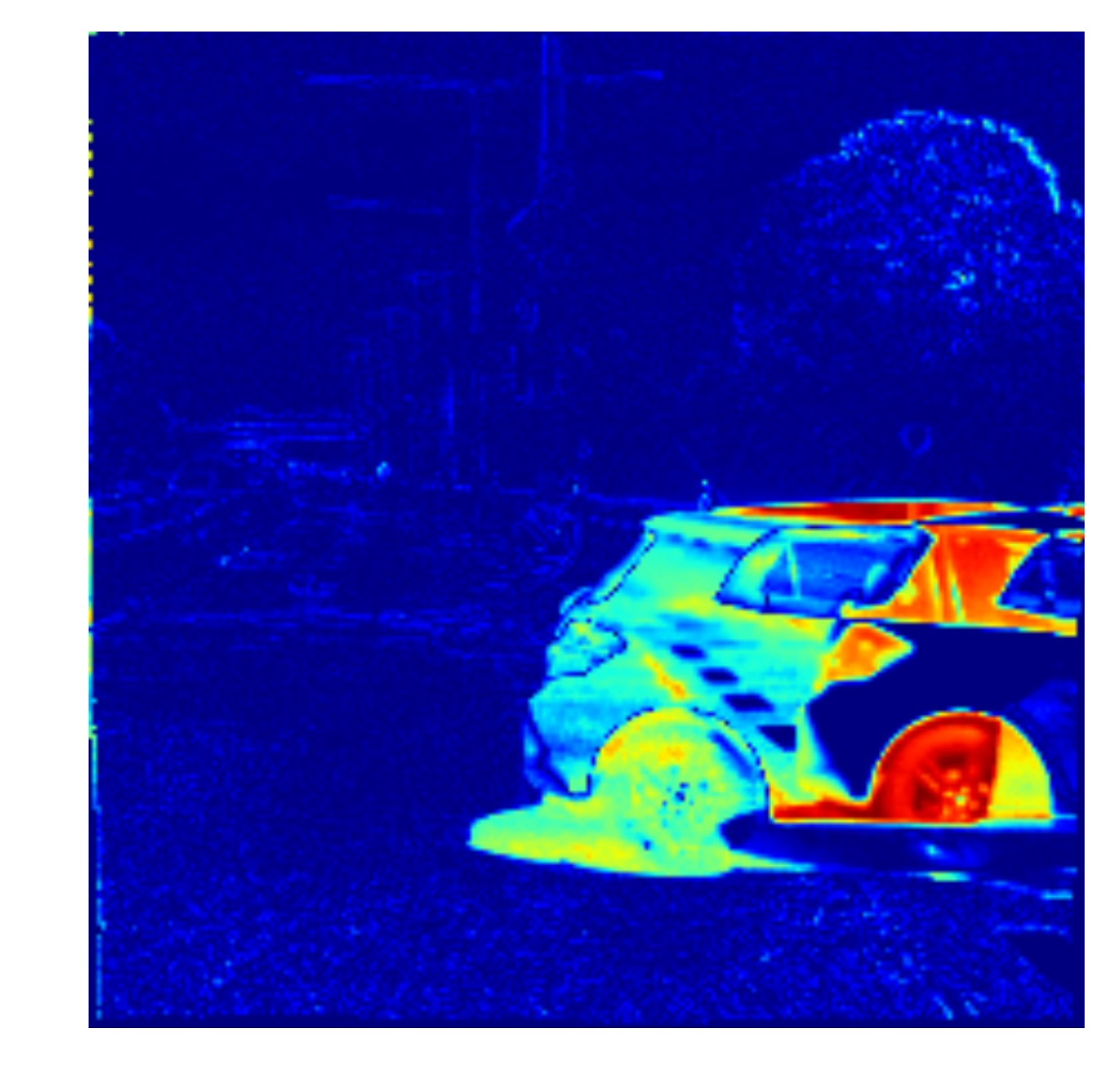}
	 	 	\end{subfigure}%
	 	 \end{subfigure}%

		\begin{subfigure}{0.33\linewidth}
			\begin{subfigure}{\linewidth}
				\includegraphics[width=\textwidth]{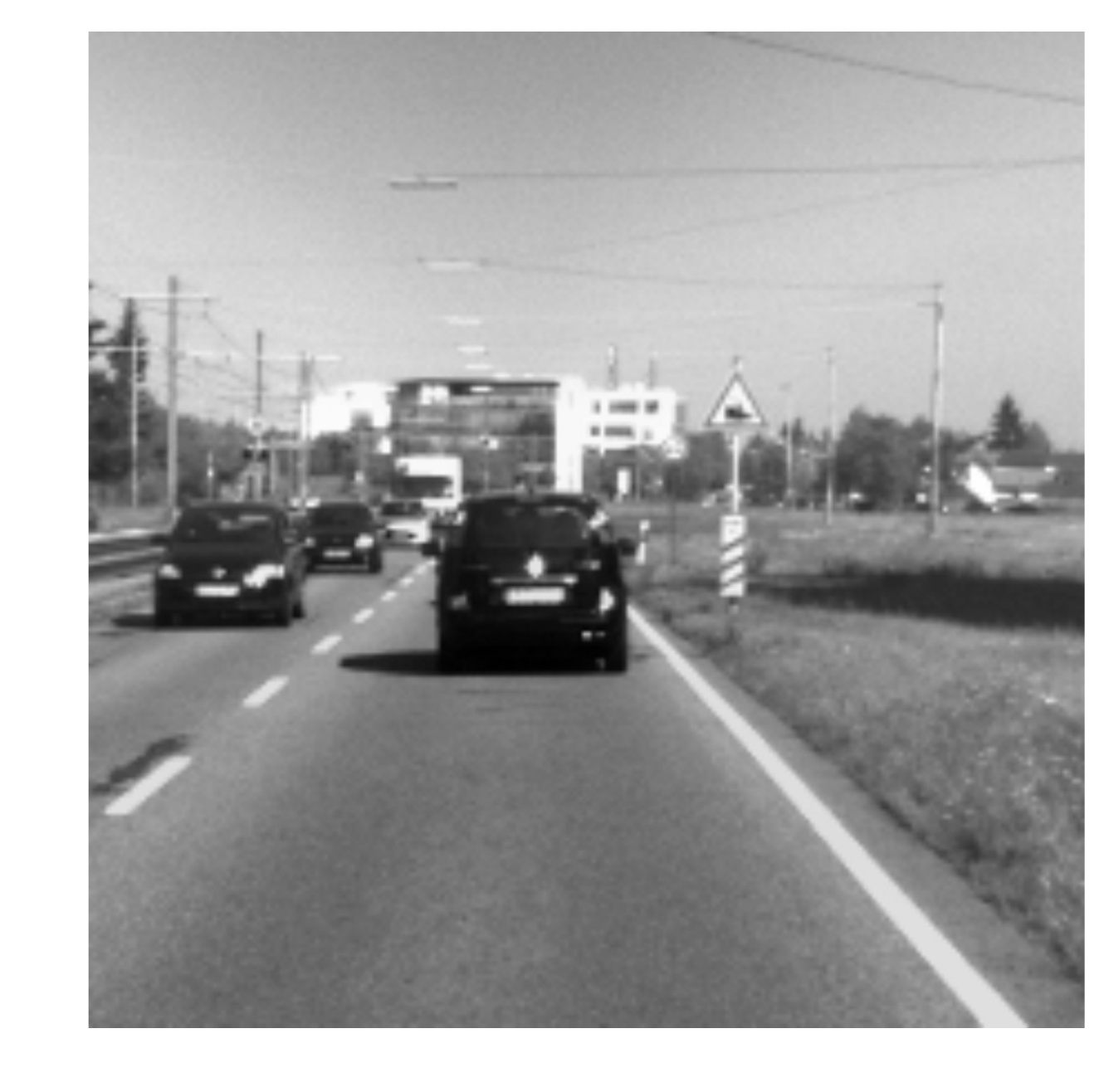}
				\caption{Source}
	  		\end{subfigure}%
			\begin{subfigure}{\linewidth}
				\includegraphics[width=\textwidth]{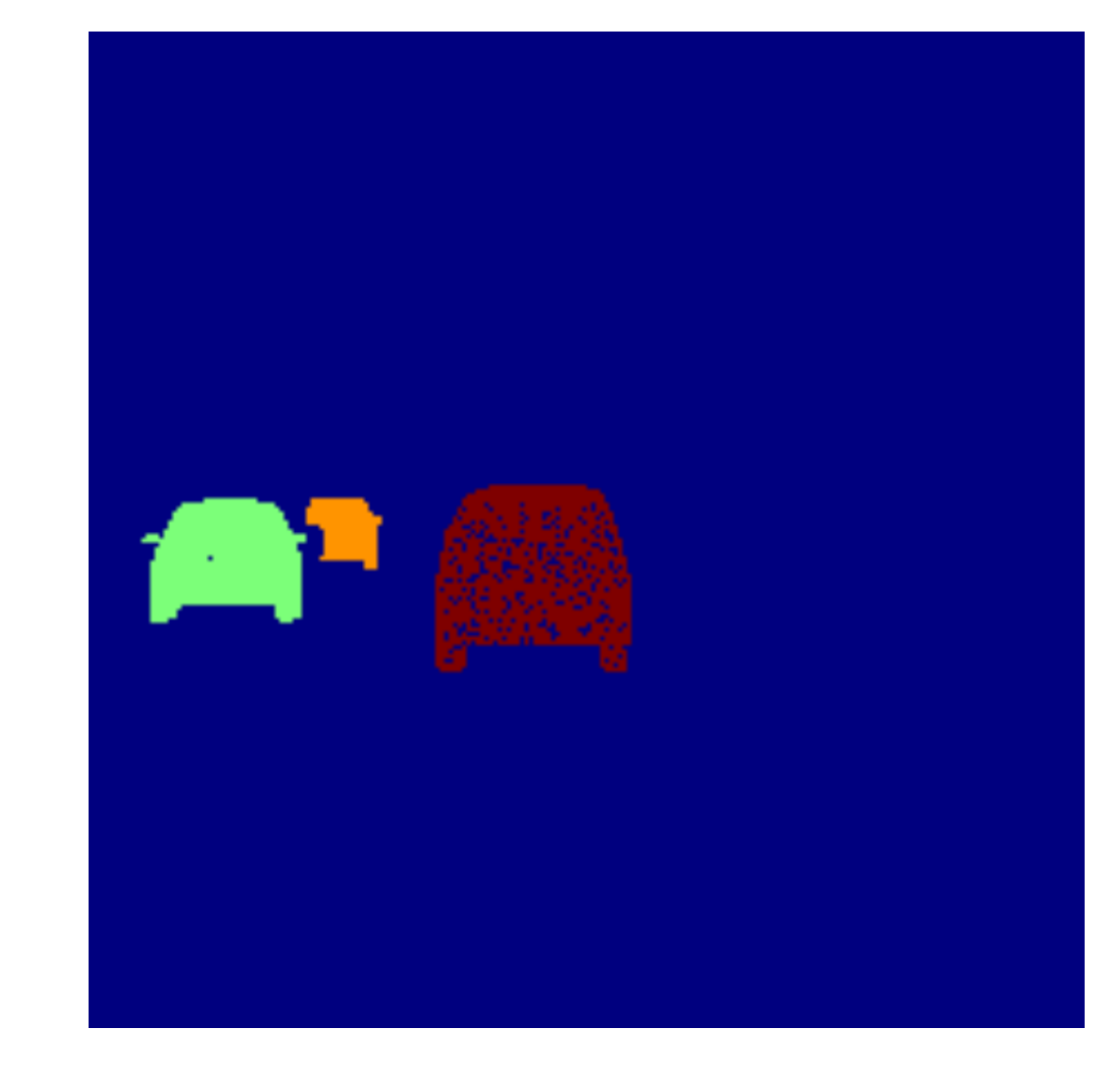}
				\caption{GT}
	  		\end{subfigure}%
			\begin{subfigure}{\linewidth}
				\includegraphics[width=\textwidth]{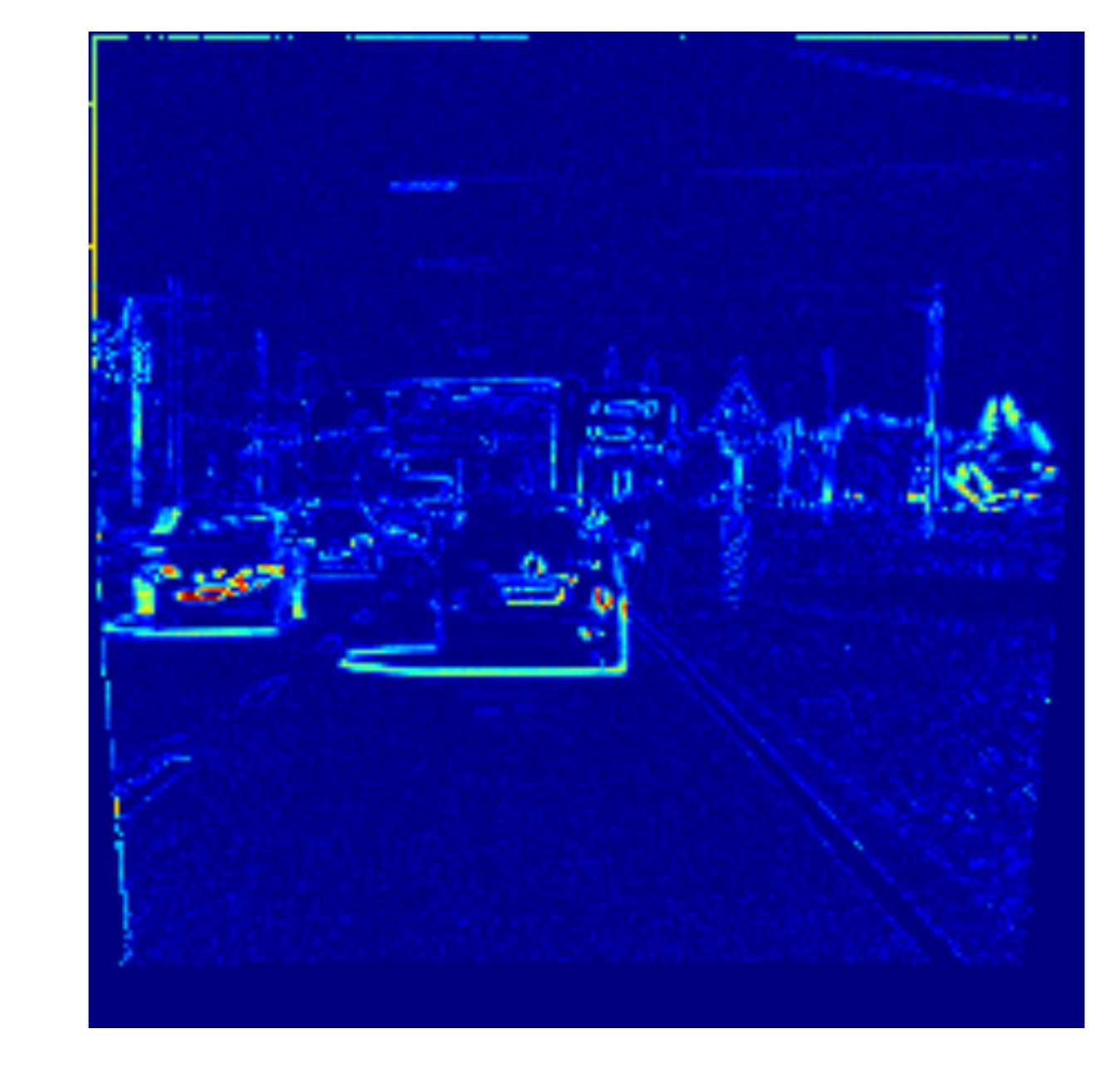}
				\caption{Residual}
	 	 	\end{subfigure}%
	 	 \end{subfigure}%
	 	 	 		  	
	 \end{minipage}
	 \end{center}
	\caption{Sample anomalous independent motion detection. Notice how  model-error induced by distant objects could potentially be confused for independent motion in the last row for the VIFlow residuals. }		
	\label{anom_detection} 
\end{figure}  

\footnotetext{Evaluated only at sparse DM keypoints (average of 676 and 565 keypoints on KITTI and EuRoC test sets, respectively)}
\subsection{Anomaly Detection} \label{section:anomaly_detection}

VIFlow also holds promise as a visual anomaly detector. Specifically, as a detector of independent motion that is inconsistent with global motion both inferred visually (during training) and the motion priors encoded by the IMU measurements and SFFMS inputs. ÒFig \ref{sample_flow}Ò presents optical flow renderings alongside ground-truth flow and flow computed by Flownet2 where ÒFig \ref{matching_results}Ò shows correspondence matchings discovered by VIFlow. As expected, the flow learned by the network is correlated with ego-motion and the network generally fails to predict large sources of independent motion.

The images shown in ÒFig. \ref{anom_detection}Ò are the residuals between the reconstructed VIFlow source outputs and the original source images (e.g. $\lVert Ir(\theta,I_{i+1}) - I_{i} \rVert$). As the network is trained to use a motion prior to help calculate visual transformations and scene flow, it is biased toward global motions and implicitly outputs mappings that correspond to ego-motion. Thus, as seen in ÒFig. \ref{anom_detection}Ò, areas of high residual magnitude can correspond to regions containing independently moving scene elements (see ÒSection \ref{anom_limitations}Ò for a discussion of what else this may correspond to).

\section{Conclusions and Future Work} \label{conclusion}

On the EuRoC MAV dataset \cite{burris2016} and on the KITTI SceneFlow 2015 dataset \cite{Geiger2013}, we have shown how an unsupervised deep network can learn to efficiently estimate visual flow from inertial measurements. The resulting runtimes for our networks are substantially faster than other SOA vision-only matching and flow algorithms with similar performance (in the case of EuRoC) to deep matching approaches.

\subsection{Limitations}

\subsubsection{Transferability and Error Correction}
Because the network receives raw IMU measurements and learns both how to integrate them and transform them to a camera reference frame, directly transferring models trained on one dataset with a given IMU and camera in some configuration to another dataset with a different IMU and a different camera in some other configuration is not currently possible. VIFlow does not receive any form IMU intrinsics or IMU/camera extrinsics and input IMU measurements are not pre-calibrated. An area for future exploration will be incorporating estimated IMU intricics (e.g., derived from the IMU datasheet), pre-calibrated IMU measurements, and explicit IMU-camera extrinsics (e.g. derived from \cite{furgale2013}) to create a parametrized version of VIFlow that can then be transferred between various datasets. However, it is unclear how  such a parametrized network will perform compared to the current iteration of VIFlow which learns to optimally integrate IMU data and transform it to the camera frame so as to minimize residual error.

Because VIFlow is a strictly feed-forward model, it is limited in its ability to correct for error online. The multi-hypothesis approach allows the network to handle certain types of structured and unstructured noise but we speculate that VIFlow may show sensitivity to shifts in the relative positioning between the IMU and camera as well as abrupt collisions. Future work will evaluate VIFlow for these sensitives and investigate online models of error correction to mitigate their effects.

\subsubsection{Anomaly Detection Vs. Model Error} \label{anom_limitations}
As presented above, one potential use for VIFlow is as a detector of anomalous independent motion. By taking the residual between a source image and a VIFlow-reconstructed source image, high-magnitude regions of interest (ROIs) can correspond to independently moving scene elements. However, these same regions can correspond to ROIs that the model failed to accurately predict (as can be seen with the distant treeline in  ÒFig. \ref{Flow_results}Ò). Another area of future work will be expanding VIFlow as an anomalous motion detector by building an additional mechanism that separates model error from independent motion-induced error. One potential path to this segmentation is to impose shape-related constraints when analyzing residuals. For example, model errors tend to be most pronounced at the edges of objects, leading to narrow bands of error in either the $X$ or $Y$ dimensions but rarely in both. Contrastingly, independent motions present with a larger spatial extent in both the $X$ and $Y$ dimensions simultaneously. Thus, detecting blobs of high-magnitude residuals and classifying them according to patterns of spatial extent may lead to the efficient separation of anomalous independent motion from model error.

\subsubsection{Surrogate Motor Signals} \label{motor_limitations}
In this work, we used a K-Means clustering of ground-truth poses to form a surrogate feed-forward motor signal as an additional input to the network. While this signal exhibited high error (see ÒFig. \ref{hist_kmeans}Ò), its error characteristics are unlikely to match that of a real-world motor input. Future work will need to examine how the network performs when input with a true feed-forward motor signal.




%

%


\bibliographystyle{IEEEtran}   			
\bibliography{IEEEabrv,hetero}  

\appendices
\section{EuRoC Ground Truth Generation} \label{appendix_euroc_gt}



To obtain depth estimates for each point in each grayscale image, we rendered range images from the ground truth point cloud of the \textit{Vicon Room}. For each image and rendered depth pair, we ray traced each pixel coordinate $(u_{t0},v_{t0})$ using the horizontal and vertical fields of view calculated from the focal lengths in the camera matrix $K$, normalized the resulting $[X,Y,1]^{T}$ coordinates, and multiplied by the depth at each pixel location to generate coordinates in the camera frame $[X_{c}^{t_{0}},Y_{c}^{t_{0}},Z_{c}^{t_{0}},1]^{T}$. 

Then, for the 4x4 transformation matrix $H_{WC}^{t_{0}}$ that transforms a vector from the camera frame $C$ to the world frame $W$ at time $t_{0}$, and another transformation matrix $H_{WC}^{t_{1}}$ that transforms a vector from the camera frame $C$ to the world frame $W$ at time $t_{1}$, we calculated the 4x4  transform matrix $H_{W_{t_0}W_{t_1}}$ as

\begin{equation}
H_{W_{t_0}W_{t_1}}={H_{WC}^{t_{0}}}^{-1}*H_{WC}^{t_{1}}
\end{equation}

\noindent and then projected points in the camera frame from $t_{0}$ to $t_{1}$:

\begin{equation}
[X_{c}^{t_{1}},Y_{c}^{t_{1}},Z_{c}^{t_{1}},1]^{T} = H_{W_{t_0}W_{t_1}}*[X_{b}^{t_{0}},Y_{b}^{t_{0}},Z_{b}^{t_{0}},1]^{T}
\end{equation}

Finally, we applied the camera matrix $K$ to project points $[X_{c}^{t_{1}},Y_{c}^{t_{1}},Z_{c}^{t_{1}},1].T$ to the imaging plane and recover ground truth-pixel coordinates:

\begin{equation}
[u_{t1},v_{t1},1]^{T} = K*[X_{c}^{t_{1}},Y_{c}^{t_{1}},Z_{c}^{t_{1}},1].T  
\end{equation}

The recovered mapping from $[u_{t0},v_{t0},1]^{T}$ to $[u_{t1},v_{t1},1]^{T}$ allows us to project points in the visual scene from one camera position to another and thus provides ground truth correspondence between two image frames.


\end{document}